\documentclass[11pt]{article}

\usepackage{graphicx}
\usepackage{amsmath}
\usepackage{amsfonts}
\usepackage{amssymb}
\usepackage{graphicx}
\usepackage{caption2}

\newcommand{\be}{\begin{equation}}
\newcommand{\ee}{\end{equation}}

\newcommand{\bea}{\begin{eqnarray}}
\newcommand{\eea}{\end{eqnarray}}

\newcommand{\bi}{\begin{itemize}}
\newcommand{\ei}{\end{itemize}}

\newcommand{\ben}{\begin{enumerate}}
\newcommand{\een}{\end{enumerate}}

\newcommand{\bef}{\begin{figure}[tbp]}
\newcommand{\enf}{\end{figure}}

\newcommand{\bt}{\begin{tabular}{lcllcl}}
\newcommand{\et}{\end{tabular}}

\newcommand{\bd}{\begin{description}}
\newcommand{\ed}{\end{description}}

\newtheorem{theorem}{Theorem}

\newtheorem{corollary}{Corollary}

\newcounter{example}

\newenvironment{proof}[1]
 {\noindent%
 {\bf \boldmath Proof #1:}}
 {\hfill $\Box$  \\}

\newcommand{\eref}[1]{(\ref{#1})}       

\newcommand{\dfn}{\stackrel{\triangle}{=}}  

\newcommand{\comb}[2]{\left (
 \raisebox{-4pt}{$\stackrel{\mbox{\large $#1$}}{#2}$} \right ) }

\newcommand{\nvec} {{\mathbf n}}

\newcommand{\pvec}   {\mbox{\boldmath $\theta$}}
\newcommand{\psivec} {\mbox{\boldmath $\psi$}}
\newcommand{\sigvec} {\mbox{\boldmath $\sigma$}}

\newcommand{\etavec} {\mbox{\boldmath $\eta$}}
\newcommand{\xivec}  {\mbox{\boldmath $\xi$}}

\begin{document}

\title{Patterns of i.i.d.\ Sequences and Their Entropy - Part II: Bounds for Some Distributions\footnote{Supported
in part by NSF Grant CCF-0347969.  Parts of the material in this
paper were presented at the IEEE International Symposium on
Information Theory, Seattle, WA, USA, July, 2006.}}
\author{Gil I. Shamir \\
 Department of Electrical and Computer Engineering \\
 University of Utah \\
 Salt Lake City, UT 84112, U.S.A \\
 e-mail: gshamir@ece.utah.edu.}
\date{}
\maketitle

\begin{abstract}

A \emph{pattern\/} of a sequence is a sequence of integer indices
with each index describing the order of first occurrence of the
respective symbol in the original sequence. In a recent paper, tight
general bounds on the block entropy of patterns of sequences
generated by independent and identically distributed (i.i.d.)\
sources were derived. In this paper, precise approximations are
provided for the pattern block entropies for patterns of sequences
generated by i.i.d.\ uniform and monotonic distributions, including
distributions over the integers, and the geometric distribution.
Numerical bounds on the pattern block entropies of these
distributions are provided even for very short blocks.
Tight bounds are obtained even for distributions that have infinite
i.i.d.\ entropy rates.  The approximations are obtained using
general bounds and their derivation techniques.  Conditional index
entropy is also studied for distributions over smaller alphabets.

\end{abstract}

\section{Introduction}
\label{sec:introduction}

Recent work (see, e.g., \cite{aberg97}, \cite{jevtic02},
\cite{orlitsky04}, \cite{shamir03},
\cite{shamir04c}, \cite{shamir04d}) has considered universal
compression for \emph{patterns\/} of independent and identically
distributed (i.i.d.)\ sequences. The pattern of a sequence
$x^n \dfn \left ( x_1, x_2, \ldots, x_n \right )$
is a sequence
$\psi^n \dfn \psivec \dfn \Psi \left ( x^n \right )$
of pointers that point to the actual alphabet letters,
where the alphabet letters are assigned \emph{indices\/} in order of
first occurrence.
For example, the pattern of all sequences $x^n = lossless$,
$x^n = sellsoll$, $x^n = 12331433$, and $x^n = 76887288$,
which is alphabet independent,
is $\psi^n =\Psi \left ( x^n \right ) = 12331433$.  Capital $\Psi (\cdot)$ denotes the pattern operator.

Patterns are interesting in universal compression with unknown alphabets,
where the dictionary and the pattern of $x^n$ can be compressed separately (see, e.g., \cite{aberg97}).
Pattern entropy is also important in
learning applications.   Consider all the new species an explorer observes.  The explorer
can identify these species with the first time each was seen, and assign
indices to species in order of first occurrence.
The entropy of patterns
can thus model uncertainty of such processes.

Initial work on patterns
\cite{jevtic02}, \cite{orlitsky04}, \cite{shamir03}, \cite{shamir04c}, \cite{shamir04d},
focused on showing diminishing universal compression redundancy rates.
The first results on pattern entropy
in \cite{shamir03}, \cite{shamir03a}, \cite{shamir04d}, however,
showed that for sufficiently large alphabets, the pattern block entropy must decrease
from the i.i.d.\ one even more significantly than the universal coding penalty for
coding patterns.
Since $\Psi \left (x^n \right )$ is the result of data processing,
its entropy must be no greater than $H_{\theta} \left ( X^n \right )$.
For alphabet size $k$,
\be
 \label{eq:entropy_bounds_simple}
 nH_{\theta} \left (X \right ) - \log \left [ k!/ \left (\max \left \{ 0, k-n \right \} \right)! \right ]  \leq
 H_{\theta} \left ( \Psi^n \right ) \leq
 nH_{\theta} \left (X \right ),
\ee
where capital letters denote random variables, and $\pvec$ is the parameter
vector governing the source\footnote{Logarithms are
taken to base $2$, here and elsewhere.  The natural logarithm is denoted by $\ln$.}.
The bounds in \eref{eq:entropy_bounds_simple} already show that
for $k = o(n)$
the pattern entropy rate equals the i.i.d.\ one
\footnote{For two functions $f(n)$ and $g(n)$, $f(n) = o(g(n))$ if
$\forall c, \exists n_0$, such that, $\forall n > n_0$, $\left |f(n)\right | <
c\left |g(n) \right |$; $f(n) = O(g(n))$ if $\exists c, n_0$, such that, $\forall n
> n_0$, $0 \leq \left |f(n)\right | \leq c\left |g(n) \right |$;
with inequalities it will
be assumed that $f(n) \geq 0$, but with equalities, negative $f(n)$ are possible;
$f(n) = \Theta(g(n))$ if $\exists c_1, c_2, n_0$,
such that, $\forall n > n_0$, $c_1 g(n) \leq f(n) \leq c_2 g(n)$.}
for non-diminishing $H_{\theta}(X)$.
Subsequently to the results in \cite{shamir03a}, it was
independently shown in \cite{gemelos06} and \cite{orlitsky06}
that for discrete i.i.d.\ sources,
the pattern entropy \emph{rate\/} is
equal to that of the underlying i.i.d.\ process.

In contrast with the block entropy,
for smaller alphabets, the conditional next index entropy $H_{\theta}\left (\Psi_{\ell} ~|~ \Psi^{\ell-1} \right )$
is guaranteed to
start increasing from $H_{\theta}(X)$ after some time $\ell >1$.
The gain (decrease) in block entropy is thus due to first occurrences of new symbols,
and gains only occur before first occurrences become sparse.
This observation, pointed out also in
\cite{gemelos06}, \cite{orlitsky06}, gives rise to a possibility of diminishing $o(1)$ overall
\emph{per block\/}
decreases of $H_{\theta} \left ( \Psi^n \right )$ from $n H_{\theta}(X)$.

In \cite{shamir07}, general
tight upper and matching lower bounds on the block entropy were derived.
This paper continues the work in \cite{shamir07}, and uses the bounds derived
in \cite{shamir07} and their derivation methods to provide very accurate approximations
of the pattern block entropies for uniform and several monotonic i.i.d.\ distributions.
The complete range of uniform distributions, from over fixed small alphabets, to over infinite
alphabets, is studied.  Monotonic distributions from slowly to fast
decaying ones are considered.  It is shown that the pattern entropy
can be approximated even for slowly decaying monotonic
distributions with infinite i.i.d.\ entropy rates.  Then, small alphabets and their
conditional next index entropies are studied.

The derivation methods are based on those in \cite{shamir07}.
The probability space is partitioned into a \emph{grid\/} of points.
Between each two points, there is a \emph{bin\/}.
Symbols whose probabilities lie in the same bin can be exchanged in
$x^n$ to provide another sequence $x'^n$ with
$\Psi \left (x^n \right ) = \Psi \left (x'^n \right )$
and almost equal probability.
Counting such sequences, packing low probability symbols into single point masses,
leads to bounds on $H_{\theta} \left ( \Psi^n \right )$.
Proper choices of grids are key for tightening bounds.

Section~\ref{sec:note_def} gives some preliminaries.
General bounds (somewhat modified) from \cite{shamir07} are reviewed in Section~\ref{sec:gen_bounds}.
Next, Section~\ref{sec:bounds} summarizes pattern entropies
for different distributions.
Finally,
Sections~\ref{sec:uniform} and~\ref{sec:monotonic}
contain the proofs
for uniform distributions and monotonic distributions, respectively.

\section{Preliminaries}
\label{sec:note_def}

Let $x^n$ be an $n$-tuple with components
$x_i \in \Sigma \dfn \left \{1, 2, \ldots, k \right \}$ (where the alphabet is defined without
loss of generality).
The asymptotic regime is that $n \rightarrow \infty$.  However, the general bounds
are stated also for finite $n$.  The alphabet size $k$ may
be greater than $n$ or infinite.
The vector $\pvec \dfn \left ( \theta_1, \theta_2, \ldots, \theta_k \right )$
is the set of probabilities of all letters in $\Sigma$.
Assume, without loss
of generality, that
$\theta_1 \leq \theta_2 \leq \cdots \leq \theta_k$.
Boldface letters denote vectors,
and capital letters will denote random variables.
The probability of $\psi^n$ \emph{induced\/} by an i.i.d.\ source is
\be
\label{eq:pattern_probability}
 P_{\theta} \left ( \psi^n \right ) =
  \sum_{y^n: \Psi (y^n) = \psi^n } P_{\theta} \left ( y^n \right ).
\ee
The \emph{pattern sequence\/} or \emph{block entropy\/} of order $n$ is
\be
\label{eq:pattern_entropy}
 H_{\theta} \left ( \Psi^n \right ) \dfn
 -\sum_{\psi^n} P_{\theta} \left ( \psi^n \right )
 \log P_{\theta} \left ( \psi^n \right ).
\ee

Following \cite{shamir07} (but more generally),
consider two different grids: $\etavec$, and $\xivec$.
For simplicity of notation, we omit the dependence on $n$ from definitions of grid points.
Let $\varepsilon_0$, $\varepsilon_1$, and $\varepsilon_2$ be three numbers that satisfy
$\varepsilon_0 \geq \max (0, \varepsilon_1)$ and $\varepsilon_2 \geq \max(0, \varepsilon_1)$.
Define
\be
 \label{eq:eta_grid_def2}
 \eta'_b \dfn
 \sum_{j=1}^{b} \frac{2 (j - \frac{1}{2})}{n^{1+\varepsilon_2}} =
 \frac{b^2}{n^{1+\varepsilon_2}}.
\ee
The grid $\etavec \dfn \left ( \eta_0, \eta_1, \ldots, \eta_{B_{\eta}} \right )$ is
defined by $\eta_0 = 0$, $\eta_1 = \frac{1}{n^{1+\varepsilon_0}}$,
$\eta_2 = \frac{1}{n^{1+\varepsilon_1}}$,
$b' \dfn b+\left \lfloor n^{(\varepsilon_2 - \varepsilon_1)/2}  \right \rfloor - 2$, and
\be
\label{eq:eta_grid_def}
 \eta_b = \eta'_{b+\left \lfloor n^{(\varepsilon_2 - \varepsilon_1)/2} \right \rfloor - 2}
 \dfn \eta'_{b'},~~
  b = 3,4, \ldots, B_{\eta}.
\ee
For some $\varepsilon > 0$, if $\varepsilon_0 = \varepsilon$, $\varepsilon_1 = -\varepsilon$,
and $\varepsilon_2 = 2\varepsilon$, $\etavec$ reduces to the one defined
in \cite{shamir07}.  The more general definition here allows
achieving tighter bounds for some specific distributions also for finite $n$.
The relation $\varepsilon_1 = -\varepsilon$ will be assumed by definition, but the other
two parameters will not be tied to $\varepsilon$ (except by $\varepsilon_b \geq -\varepsilon$).
The grid $\xivec \dfn \left ( \xi_0, \xi_1, \ldots, \xi_{B_{\xi}} \right )$ is defined by
$\xi_0 = 0$, and for an arbitrarily small $\varepsilon = -\varepsilon_1 > 0$,
\be
 \label{eq:xi_grid_def}
 \xi_b \dfn \sum_{j=1}^b \frac{2(j-0.5)}{n^{1-\varepsilon}} =
 \frac{b^2}{n^{1-\varepsilon}},~~ b = 1, 2, \ldots, B_{\xi}.
\ee
For both grids, $\eta_{B_{\eta}+1} = \xi_{B_{\xi}+1} \dfn 1$, and thus
$B_{\eta} = \left \lfloor \sqrt{n}^{1+\varepsilon_2} \right \rfloor -
\left \lfloor n^{(\varepsilon_2-\varepsilon_1)/2} \right \rfloor + 2$, and
$B_{\xi} = \left \lfloor \sqrt{n}^{1-\varepsilon} \right \rfloor$.
We also define the maximal indices $A_{\eta}$, and $A_{\xi}$ whose
grid points do not exceed $0.5$ for $\etavec$, and $\xivec$, respectively.
Hence,
$A_{\eta} = \left \lfloor \sqrt{n}^{1+\varepsilon_2}/ \sqrt{2} \right \rfloor -
\left \lfloor n^{(\varepsilon_2-\varepsilon_1)/2} \right \rfloor + 2$, and
$A_{\xi} = \left \lfloor \sqrt{n}^{1-\varepsilon} /\sqrt{2} \right \rfloor$.

Let $k_b$, $b=0, 1, \ldots, B_{\eta}$; and $\kappa_b$,
$b=0, 1, \ldots, B_{\xi}$; denote the numbers of symbols
$\theta_i \in \left (\eta_b, \eta_{b+1} \right ]$,
and
$\theta_i \in \left (\xi_b, \xi_{b+1} \right ]$, respectively (in bin $b$ of
$\etavec$ and $\xivec$, respectively).
Specifically, for given $\varepsilon_0$, $\varepsilon$, and $\varepsilon_2$,
$k_0$ and $k_1$ denote the cardinalities of
$\theta_i \leq 1/n^{1+\varepsilon_0}$, and
$\theta_i \in \left ( 1/n^{1+\varepsilon_0}, 1/n^{1-\varepsilon} \right ]$,
respectively.
Define also $k_{01} \dfn k_0 + k_1$, thus $k-k_{01}$ is the cardinality of
$\theta_i > 1/n^{1-\varepsilon}$.
Also, let $\kappa'_b$, $b = 1, 2,\ldots, B_{\xi}$; be zero
if $\kappa_b$ is zero, and otherwise, the number of symbols for which
$\theta_i \in \left (\xi_{b-1}, \xi_{b+2} \right ]$,
with the exception of $\kappa'_1$,
which will only count letters for which
$\theta_i \in \left (\xi_1, \xi_3 \right ]$.
(There is clearly
an overlap between adjacent counters $\kappa'_b$, which is needed for one
of the lower bounds.)
Now, let
\be
 \varphi_b \dfn \sum_{\theta_i \in \left (\eta_b, \eta_{b+1} \right ]} \theta_i
\ee
be the total probability of bin $b$ of
$\etavec$.
Specifically, $\varphi_0$,
$\varphi_1$, and $\varphi_{01} \dfn \varphi_0 + \varphi_1$ are
defined with respect to (w.r.t.)\ bins $0$, $1$, and $01$, respectively.

The mean occurrence count of letter $i$ in $X^n$ is given by
$E_{\theta} N_x \left ( i \right ) = n \theta_i$,
where $n_x \left ( i \right )$ is the occurrence count of $i$ in $x^n$,
$N_x \left ( i \right )$ is its random variable, and $E_{\theta}$ is expectation
given $\pvec$.
Let $K_b$ be a random variable counting the \emph{distinct\/}
symbols from bin $b$ of $\etavec$
that occur in $X^n$.
Let $K$ be the total distinct letters occurring in $X^n$.
Then, let
\be
 \label{eq:mean_bin}
 L_b \dfn E_{\theta} \left [ K_b \right ]=
 \sum_{i~:~\theta_i \in \left ( \eta_b, \eta_{b+1} \right ]}
 P_{\theta} \left ( i \in X^n \right ) =
 \sum_{\theta_i \in \left ( \eta_b, \eta_{b+1} \right ]}
 \left [ 1 - \left ( 1 - \theta_i \right )^n \right ]
\ee
and also define $L \dfn E_{\theta} \left [ K \right ]$ similarly.
Substituting $(1-\theta_i)^n = \exp \left \{ n \ln (1-\theta_i) \right \}$ and using Taylor series expansion,
\be
 \label{eq:mean_bin_bound}
 k_b - \sum_{\theta_i \in \left ( \eta_b, \eta_{b+1} \right ]}
 e^{-n \theta_i} \leq
 L_b \leq
 k_b - \sum_{\theta_i \in \left ( \eta_b, \eta_{b+1} \right ], ~\theta_i \leq 3/5}
  e^{-n \left (\theta_i+\theta_i^2 \right )}.
\ee
Specifically, using Binomial expansion for bin $b=0$,
\be
 \label{eq:min_bin0_bound}
 n \varphi_0 - \comb{n}{2} \sum_{i=1}^{k_0}
 \theta_i^2 \leq L_0 \leq
 n \varphi_0 - \comb{n}{2} \sum_{i=1}^{k_0}\theta_i^2 +
 \comb{n}{3} \sum_{i=1}^{k_0}\theta_i^3.
\ee
Similar bounds can be obtained for bin $b=1$ if $\varepsilon_1 \geq 0$ ($\varepsilon \leq 0$).

\section{General Bounds}
\label{sec:gen_bounds}

General bounds based on \cite{shamir07} are summarized here.  First, for
given $\varepsilon_0$, $\varepsilon$, and $\varepsilon_2$, that determine
$\etavec$ and $\xivec$, define
\bea
 \label{eq:zeroone_bin_packed_entropy}
 H_{\theta}^{(01)} \left ( X \right ) &\dfn&
 -\varphi_{01} \log \varphi_{01} - \sum_{i=k_{01}+1}^k \theta_i \log \theta_i, \\
 \label{eq:zero_one_bin_packed_entropy}
 H_{\theta}^{(0,1)} \left ( X \right ) &\dfn&
 -\sum_{b=0}^1 \varphi_b \log \varphi_b -
 \sum_{i=k_{01}+1}^k \theta_i \log \theta_i.
\eea
The i.i.d.\ entropies above pack low probabilities into one or two point masses.
The following lower bounds were derived in
\cite{shamir07}.
\begin{theorem}
\label{theorem:lb}
Let $\varepsilon > 0, \varepsilon_0 \geq 0$, and define
$\xivec$ with \eref{eq:xi_grid_def}.
Define $Z^n \dfn \left (Z_1, Z_2, \ldots, Z_n \right )$ by
$Z_j = 0$ if $\theta_{X_j} \leq 1/n^{1-\varepsilon}$, and $1$ otherwise.
Let $k_{\vartheta}^-$ be the count of letters $i$ such that
$\theta_i \in \left ( \vartheta^- / n^{1-\varepsilon}, 1/ n^{1-\varepsilon} \right ]$ and
$k_{\vartheta}^+$ the count of letters $i$ with
$\theta_i \in \left ( 1/ n^{1-\varepsilon}, \vartheta^+/ n^{1-\varepsilon} \right ]$,
where $\vartheta^-$, $\vartheta^+$ are constants that satisfy $\vartheta^+ > 1 > \vartheta^- > 0$.
Then,
\be
 \label{eq:lb2}
 H_{\theta} \left ( \Psi^n \right ) \geq
 n H^{(01)}_{\theta} \left ( X \right ) - S_1 + S_2 + S_3 - S_4
\ee
where
\bea
 \label{eq:lb2_S1b0}
 S_1 &\leq& \log (k - k_{01} )! \\
 \label{eq:lb2_S1b1}
 S_1 &\leq&
 \left (1 - \varepsilon_n \right ) \left \{
 \sum_{b=1}^{A_{\xi}} \log \left ( \kappa_b ! \right ) + \left (k - k_{01} \right ) \log 3 \right \} +
 \varepsilon_n \log ( k - k_{01} )! + h_2 \left [ \min \left ( \varepsilon_n, 0.5 \right ) \right ]\\
 \label{eq:lb2_S1b2}
 S_1 &\leq& \left ( 1 - \varepsilon_n \right ) \sum_{b=1}^{A_{\xi}} \log \left ( \kappa'_b ! \right ) +
 \varepsilon_n \log ( k - k_{01} )! + h_2 \left [ \min \left ( \varepsilon_n, 0.5 \right ) \right ]\\
 \label{eq:epsin_def}
 \varepsilon_n &\dfn& \min \left \{n \cdot (k - k_{01}) \cdot e^{-0.1n^{\varepsilon}}, 1 \right \}
\eea
where $h_2(\alpha) \dfn -\alpha \log \alpha - (1-\alpha) \log (1 - \alpha)$,
\bea
 \label{eq:lb2_S2b0}
 S_2 &=&
 \sum_{i=1}^{k_{01}} E_{\theta}
 \left [ N_x(i) - P_{\theta} \left ( i \in X^n \right ) \right ]
 \log \frac{\varphi_{01}}{\theta_i} \\
 \label{eq:lb2_S2b1}
 S_2 &\geq&
 \sum_{i=1}^{k_{01}} \left [ n \theta_i - 1 + e^{-n \left ( \theta_i +
 \theta_i^2 \right )} \right ]
 \log \frac{\varphi_{01}}{\theta_i} \\
 \label{eq:lb2_S2b2}
 S_2 &\geq&
 \left ( 1 - \frac{1}{3n^{\varepsilon_0}} - \frac{2}{n} \right )
 \frac{n^2}{2} \sum_{i=1}^{k_0} \theta_i^2 \log \frac{\varphi_{01}}{\theta_i} +
 \sum_{i=k_0 + 1}^{k_{01}} \left [ n \theta_i - 1 + e^{-n \left ( \theta_i +
 \theta_i^2 \right )} \right ]
 \log \frac{\varphi_{01}}{\theta_i}
\eea
\be
 \label{eq:lb2_S3b}
 S_3 \geq
 (\log e) \sum_{i=1}^{L_{01}-1} \left ( L_{01} - i \right )
 \frac{\theta_i}{\varphi_{01}}
\ee
and
\be
 \label{eq:lb2_S4}
 S_4 \leq \min \left \{n h_2 \left ( \varphi_{01} \right ), ~
 \left ( 1 - \varepsilon'_n \right )
 \log \comb{k_{\vartheta}^- + k_{\vartheta}^+}{k_{\vartheta}^+} +
 \varepsilon'_n n +
 h_2 \left [ \min \left ( \varepsilon'_n, 0.5 \right ) \right ] \right \}
\ee
where
\be
 \label{eq:epsipn_def}
 \varepsilon'_n \dfn
 n \cdot k_{\theta_i > n^{-3}} \cdot
 e^{-f \left ( \vartheta^-, \vartheta^+ \right ) n^{\varepsilon}} +
 \frac{\ln \vartheta^-}{2 (\vartheta^- - 1) n^{1+\varepsilon}},
\ee
$k_{\theta_i > n^{-3}}$ denotes the total symbol count with $\theta_i > 1/n^3$, and
\be
 \label{eq:f_def}
 f \left ( \vartheta^-, \vartheta^+ \right ) \dfn
 \min \left \{
 \frac{\vartheta^{\pm} - 1}{\ln \vartheta^{\pm}}
 \ln \frac{\vartheta^{\pm} - 1}{e \cdot \ln \vartheta^{\pm}} + 1
 \right \}
\ee
where the minimum is taken between the values of the expression for
$\vartheta^-$ and for $\vartheta^+$.\\
Fix $\delta > 0$,
let $n \rightarrow \infty$, and $\varepsilon \geq (1+\delta)(\ln \ln n)/(\ln n)$.
Then, $\varepsilon_n = o(1)$, $\varepsilon'_n = o(1)$, and all terms
but the leading ones in \eref{eq:lb2_S1b1}, \eref{eq:lb2_S1b2} and in the second argument
of the minimum in \eref{eq:lb2_S4} are $o(1)$.
\end{theorem}
Second order terms are described in
Theorem~\ref{theorem:lb} more explicitly than in \cite{shamir07} and some terms are
tightened (in second order) to allow use of
the theorem for practical $n$ in Section~\ref{sec:monotonic}
(derivations of the explicit terms do follow \cite{shamir07}).
This is specifically for cases
where very slow rates are obtained for the gaps between the i.i.d.\ and pattern entropies,
such as the geometric distribution.
Term $S_1$ is the decrease in $H_{\theta} \left ( \Psi^n \right )$ due to first
occurrences of symbols with $\theta_i > 1/n^{1-\varepsilon}$, which results from
indistinguishability among indices of letters in the same bin $b \geq 1$ of $\xivec$.
Term $S_2$ is the cost of re-occurrences of letters with ``small'' probabilities.
Term $S_3$ is the penalty in first occurrences
of ``small'' probability symbols beyond a single point mass.  The bound in
\eref{eq:lb2_S3b} is under a worst case assumption.
Term $S_4$ is a correction from separation between ``small'' and ``large'' probabilities.
Specifically, for $\vartheta^- = e^{-5.5} \approx 0.004$
and $\vartheta^+ = e^{1.4} \approx 4.06$,
$f \left ( \vartheta^-, \vartheta^+ \right ) > 0.5$, and the
last term of \eref{eq:epsipn_def} is upper bounded by
$2.77/n^{1+\varepsilon}$.

The following upper bounds generalize the derivations in \cite{shamir07}:
\begin{theorem}
\label{theorem:ub3}
Let $\varepsilon_0$, $\varepsilon_1$, $\varepsilon_2$, $\etavec$, $b$, and $b'$ be as in
\eref{eq:eta_grid_def2}-\eref{eq:eta_grid_def}, and let $\varepsilon = -\varepsilon_1$. Then,
\bea
 \label{eq:ub3}
 H_{\theta} \left ( \Psi^n \right ) &\leq&
 n H^{(0,1)}_{\theta} \left ( X \right ) - U + R'_1 + R'_0 \\
 \label{eq:ub3_c1}
 H_{\theta} \left ( \Psi^n \right ) &\leq&
 n H^{(01)}_{\theta} \left ( X \right ) - U + R'_{01}
\eea
where $U \geq 0$, and also
\bea
 \nonumber
 U &\geq& \sum_{b=2}^{A_{\eta}}
 \max \left \{0,~ L_b \log \frac{L_b}{e},~
 \left ( 1 - \min \left \{1, k_b e^{-n\eta_b} \right \} \right ) \log (k_b!) \right \} \\
 \label{eq:ub3_U}
 & & -~
 \frac{\left(2+1/b' \right )^2 \log e}{n^{\varepsilon_2}}
 \sum_{b\geq 2, k_b > 1} k_b
\eea
\be
 \label{eq:ub3_Rb}
 R'_b \leq
 \left ( n \varphi_b - L_b \right ) \log \left [\min \left \{k_b, n \right \} \right ] +
 n \varphi_b \cdot h_2 \left ( \frac{L_b}{n\varphi_b} \right ), ~~b = 0, 1, 01
\ee
where \eref{eq:ub3_Rb} decreases with $L_b$ for
$b = 0$ and also for $b=1, 01$ if either $\varepsilon \leq 0$ or
$k_1, k_{01} \geq \left ( 1 + \delta \right ) n^{\delta}$ for some $\delta > 0$,
respectively. Also,
\be
 \label{eq:ub3_R0}
 R'_b \leq
 \left ( \frac{n^2}{2} \sum_{i: \theta_i \in (\eta_b, \eta_{b+1}]} \theta_i^2 \right )
 \log \frac{2 e \cdot \varphi_b \cdot \min \left \{k_b, n \right \}}
 {n \sum_{j: \theta_j \in (\eta_b, \eta_{b+1}]} \theta_j^2}
\ee
for $b=0$, and also for $b=1$, and $b=01$ if $\varepsilon \leq 0$, where
$\eta_{01} \dfn \eta_0 = 0$, and $\eta_{01+1} \dfn \eta_2$. \\
Fix $\delta > 0$,
let $n \rightarrow \infty$, and $\varepsilon, \varepsilon_2 \geq (1+\delta)(\ln \ln n)/(\ln n)$.
Then, \eref{eq:ub3_U} is
\be
 \label{eq:ub3_Us}
  U \geq \left ( 1 - o(1) \right ) \sum_{b=2}^{A_{\eta}} \log \left (k_b ! \right ).
\ee
\end{theorem}
The bounds of \eref{eq:ub3}-\eref{eq:ub3_c1} consist of 1) an i.i.d.\ entropy
which packs low probabilities into one or two point masses,
2) a correction term $U$, expressing the gain in first occurrences of symbols with $\theta_i > 1/n^{1-\varepsilon}$,
3) losses in packing low probabilities into single point masses ($R'_b$ terms).
Theorem~\ref{theorem:ub3} compacts the representation of several bounds in \cite{shamir07} by allowing
negative $\varepsilon$.  This also generalizes the upper bounds in \cite{shamir07} because
two \emph{separate\/} bins with probabilities asymptotically smaller than $1/n$ can be created.
This is useful in obtaining tighter bounds for fast decaying
distributions, such as geometric distributions (see Section~\ref{sec:monotonic}).
The proof of the generalization is identical to the proof in \cite{shamir07}.
Probability is sequentially assigned to the joint index-bin sequence $\left (\psi^n, \beta^n \right )$.
Repetitions are assigned the mean bin probability, and first occurrences of an index in a bin are
assigned the remaining bin probability.  In bins $0$ and $1$ (or bin $01$), repetitions are assigned
smaller probabilities (which are optimized), and first occurrences thus greater remaining bin probability.
The average description length of this code bounds the pattern block entropy (see \cite{shamir07} for details).
The bound in \eref{eq:ub3_U} uses the better decrease in the pattern entropy that can be obtained in
each large probability bin.  The second (second order) term is the quantization cost in all bins.
The coefficient is tightened from \cite{shamir07} based on (9) in \cite{shamir07} to allow tighter
bounds for finite $n$.

\section{Bounds for Some Distributions}
\label{sec:bounds}

\subsection{Uniform Distributions}

The pattern entropy is bounded below for the complete range of uniform distributions.
Applications, as compression with words as the single alphabet unit, can have alphabets
of $k = O(n)$ or larger.
Pattern entropy for uniform distributions with $k=O(n)$ or larger is also interesting in applications
of population estimation from limited observations (see, e.g., \cite{orlitsky07}).
For uniform distributions, all symbol probabilities are in the same bin,
(also, unlike other cases, $H_{\theta} \left ( \Psi_{\ell} ~|~ \Psi^{\ell-1} \right ) = H_{\theta}
\left ( \Psi_{\ell} ~|~ X^{\ell-1} \right )$).
This guarantees
a maximal decrease of the pattern entropy from the i.i.d.\ one for alphabets of $k=o(n)$.
For alphabets of $k=O(n)$, the analysis in \cite{shamir07}
can be simplified to derive
tighter bounds due to the simplicity of the uniform distribution.
First, however, the
bounds derived from the general bounds in Section~\ref{sec:gen_bounds}
are given in the following
corollary:
\begin{corollary}
 \label{cor:uniform}
 Let $\theta_i = \theta \geq 1/n^{1-\varepsilon}$, for  $i = 1, 2, \ldots, 1/\theta = k$.  Then,
 \be
  \label{eq:uniform_large_prob}
  nH_{\theta}(X) - \log (k!) \leq
  H_{\theta} \left ( \Psi^n \right ) \leq
  nH_{\theta}(X) - \left ( 1 - k e^{-n/k} \right ) \log (k!).
 \ee
 Let $\theta_i = \lambda/n$, for $i = 1, 2, \ldots, n/\lambda = k$,
 and a fixed $\lambda > 0$.  Then,
 \bea
  \nonumber
  \lefteqn{
  \left ( 1 - \frac{1 - e^{-\lambda}}{\lambda} \right ) n \log \frac{n}{\lambda} +
  \frac{\log e }{2} \cdot \frac{\left ( 1 - e^{-\lambda} \right )^2}{\lambda} \cdot n -
  O \left ( \log n \right ) \leq } \\
  H_{\theta} \left ( \Psi^n \right )
  \label{eq:uniform_med_prob}
  &\leq&
  \left ( 1 - \frac{1 - e^{-\lambda}}{\lambda} \right ) n
  \log \left [ \min \left \{n, \frac{n}{\lambda} \right \} \right ] +
  n \cdot h_2 \left ( \frac{1 - e^{-\lambda}}{\lambda} \right ).
 \eea
 Let $\theta_i = 1/n^{\mu + \varepsilon}$, for $i = 1, 2, \ldots, n^{\mu+\varepsilon} = k$,
 and $\mu \geq 1$.  Then,
 \be
  \label{eq:uniform_small_prob}
  \left ( 1 - O \left (\frac{1}{n^{\mu-1+\varepsilon}} + \frac{1}{n} \right ) \right )
  \frac{n^{2-\mu-\varepsilon}}{2} \log \left (e n^{\mu+\varepsilon}\right ) \leq
  H_{\theta} \left ( \Psi^n \right ) \leq
  \frac{n^{2-\mu-\varepsilon}}{2} \log \left (2 e n^{\mu+\varepsilon}\right ).
 \ee
\end{corollary}

Corollary~\ref{cor:uniform} shows the decrease of the block entropy for a uniform
i.i.d.\ distribution from the original process to its pattern.  While the i.i.d.\
entropy is always $n \log k$, the pattern entropy behaves differently in three regions.
For small $k = o(n)$, the
decrease in the block entropy is only in the second order
essentially by $\log (k/e)$ bits per probability parameter.
In the other extreme $k \gg n$, the block entropy decreases in its first order \emph{rate\/}
by a factor of $2n^{\mu - 1 +\varepsilon}$
from the i.i.d.\ one.  If $\mu \geq 2$, while both the i.i.d.\ entropy
rate and block entropy diverge, the pattern entropy for the whole
block diminishes.  This is expected since for such distributions the only
pattern one expects to observe is $\psi^n = 123\ldots n$.
In the middle range
($k = O(n)$), the decrease is in the first order \emph{coefficient\/}.
Specifically,
for $\lambda = 1$, the bounds in \eref{eq:uniform_med_prob} reduce to
\be
\label{eq:uniform1n}
 \frac{n}{e} \log n + \frac{\log e}{2} \left ( 1- \frac{1}{e} \right )^2 \cdot n-
 O \left ( \log n \right ) \leq
 H_{\theta} \left ( \Psi^n \right ) \leq
 \frac{n}{e} \log n + n \cdot h_2 \left ( \frac{1}{e} \right ),
\ee
which yield
\be
 \label{eq:uniform1n1}
 \frac{n}{e} \log n + 0.29 n - O \left ( \log n \right ) \leq
 H_{\theta} \left ( \Psi^n \right ) \leq
 \frac{n}{e} \log n + 0.95 n.
\ee
Thus, the first order
gain (decrease) from the i.i.d.\ entropy is $\left (1 - \frac{1}{e} \right ) n \log n$ bits.
The decrease is because not all letters occur in a sequence.  The gain thus results from
higher probabilities of
occurrence of new indices.
The gaps between the lower and upper bounds in \eref{eq:uniform_med_prob} and in
\eref{eq:uniform1n}-\eref{eq:uniform1n1}
affect only second order terms.  However, tighter bounds for the middle range
of uniform distributions are possible.  Due to the simplicity of the uniform
distribution, some
looser bounding steps that are necessary to produce general bounds can be avoided.
Theorem~\ref{theorem_1n_bounds}
provides tighter bounds for the $\lambda/n$ uniform distribution.
\begin{theorem}
\label{theorem_1n_bounds}
 Let $\theta_i = \lambda/n$, for $i = 1, 2, \ldots, n/\lambda = k$,
 and a fixed $\lambda > 0$.  Then,
 \bea
  \nonumber
  \lefteqn{
  \left ( 1 - \frac{1 - e^{-\lambda}}{\lambda} \right ) n \log \frac{n}{\lambda} +
  \frac{\left (e^{\lambda} - \lambda - 1\right ) \log e}{\lambda e^{\lambda}} \cdot n -
  O \left ( \log n \right ) \leq } \\
  H_{\theta} \left ( \Psi^n \right )
  \nonumber
  &\leq&
  \left ( 1 - \frac{1 - e^{-\lambda}}{\lambda} \right ) n
  \log \left [ \min \left \{ n, \frac{n}{\lambda} \right \} \right ] +
  \frac{\left ( 1-e^{-\lambda} \right ) \log e}{\lambda} \cdot n + \\
  \nonumber
  & &
  \left \{ \log \alpha + \frac{\alpha -1}{\max (1, \lambda)} \log \left ( \alpha - 1 \right ) -
  \left ( \frac{\alpha - 1}{\max(1,\lambda)} + \frac{1 - e^{-\lambda}}{\lambda} \right )\right . \\
  \label{eq:uniform_med_prob1}
  & & \left . \cdot
  \log \left [ \left ( \alpha -1 \right ) + \frac{\max(1,\lambda)}{\lambda} \left (1 - e^{-\lambda} \right ) \right ]
  \right \} \cdot n +
  O \left ( \log n \right ),
 \eea
 where $\alpha \geq 1$ is a parameter which can be optimized to minimize the upper bound.
\end{theorem}
\bef
\centerline{\includegraphics
[bbllx=50pt,bblly=185pt,bburx=560pt,
bbury=598pt,height=8cm, width=9cm, clip=]{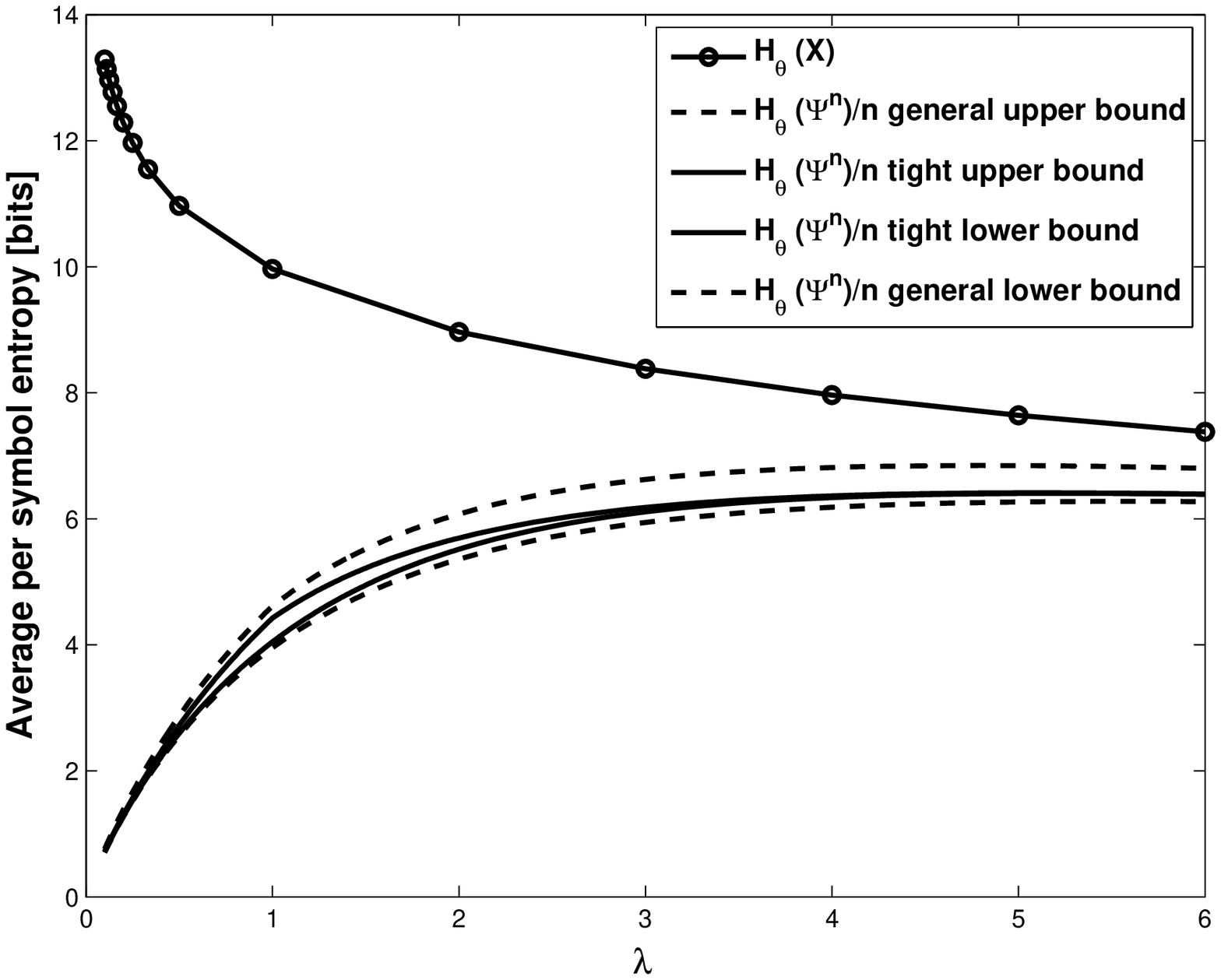}}
\caption{Bounds on pattern per symbol entropy $H_{\theta} \left ( \Psi^n \right )/n$ vs.\
$\lambda$ for uniform distributions with $\theta_i = \lambda/n, \forall i$ for
$n = 1000$ symbols.}
\label{fig:uniform_dist}
\enf
The bounds of \eref{eq:uniform_med_prob1} are tighter than those of
\eref{eq:uniform_med_prob}.
For a specific $\lambda$, the upper bound in
\eref{eq:uniform_med_prob1} is optimized by taking
$\alpha \geq 1$ that
gives a minimum.  Specifically, for $\lambda = 1$,
\be
 \label{eq:uniform1n2}
 \frac{n}{e} \log n  + 0.38n - O \left ( \log n \right ) \leq
 H_{\theta} \left (\Psi^n \right ) \leq
 \frac{n}{e} \log n + 0.76 n + O \left ( \log n \right ),
\ee
where the best choice of $\alpha$ in \eref{eq:uniform_med_prob1} leading to
\eref{eq:uniform1n2} is $\alpha \approx 1.93$.  In general, the smaller is $\lambda$, the greater
the optimal $\alpha$.  Figure~\ref{fig:uniform_dist} shows the bounds of
\eref{eq:uniform_med_prob} and \eref{eq:uniform_med_prob1} on $H_{\theta} \left ( \Psi^n \right )$
as function of $\lambda$.  It demonstrates the gaps between the i.i.d.\ block entropy and the
pattern entropy, which significantly increase the greater the alphabet is.  The bounds of
\eref{eq:uniform_med_prob1} almost meet for larger $\lambda$.

\subsection{Monotonic Distributions}

While there exist processes
for which the i.i.d.\ entropy cannot be bounded, the pattern block entropy, while
it still increases with $n$ (giving an infinite entropy rate), can be explicitly
bounded.

\subsubsection{Slowly Decaying Distribution Over the Integers}

Consider the distribution over the integers
\be
\label{eq:integer_dist}
 \tilde{\theta}_j = \frac{\alpha}{j \left ( \log j \right )^{1+\gamma}}, ~j=2,3,\ldots,
\ee
where $\gamma > 0$ and $\alpha$ is a normalizing factor.  Approximating $\sum \tilde{\theta}_j = 1$ by
integrals
\be
 \label{eq:int_alpha}
 \frac{1}{0.5 + \frac{1}{3 (\log 3)^{1+\gamma}} +
 \frac{\ln 2}{\gamma (\log 3)^\gamma}} \leq
 \alpha \leq
 \frac{1}{0.5 + \frac{\ln 2}{\gamma (\log 3)^{\gamma}}}.
\ee
The distribution in \eref{eq:integer_dist} is particularly
interesting for $0 < \gamma \leq 1$, where $H_{\theta}(X) = \infty$.
This was used to demonstrate several points in \cite{gemelos06}, \cite{orlitsky06}.   In particular,
in \cite{gemelos06}, it was used to show that there exist i.i.d.\ pattern processes with entropy
whose order is greater than $\Theta \left (n (\log n)^{1-\delta}\right )$ for every $\delta$; $0 < \delta < 1$.
Here, tight bounds approximate $H_{\theta} \left ( \Psi^n \right )$ for the
distribution in \eref{eq:integer_dist} for every $\gamma > 0$, even for relatively small $n$.
While $H_{\theta}(X) = \infty$ for $0 < \gamma \leq 1$,
for $\gamma > 1$, it is computed by
\be
 \label{eq:int_ent_sums}
 H_{\theta} \left (X \right ) = - \log \alpha +
 \sum_{j=2}^{\infty} \frac{\alpha}{j (\log j)^{\gamma}} +
 \sum_{j=3}^{\infty} \frac{\alpha (1 + \gamma) \log (\log j)}
 {j (\log j)^{1+\gamma}}.
\ee
Lower bounding the two sums by integrals
\bea
 \label{eq:int_ent_lower}
 H_{\theta} (X) &\geq&
 - \log \alpha +
 \frac{\alpha \ln 2}{\gamma - 1} +
 \frac{\alpha (1+\gamma)}{\gamma^2 (\log 3)^{\gamma}}
 \left ( 1 + \gamma \ln (\log 3) \right )
 \dfn \underline{H}_{\theta}(X), \\
 \label{eq:int_ent_upper}
 H_{\theta}(X) &\leq& \underline{H}_{\theta}(X) + \frac{\alpha}{2} +
 \frac{\alpha (1+\gamma) \log (\log 3)}{3 (\log 3)^{1+\gamma}}.
\eea
Tighter bounds (on both $\alpha$ and $H_{\theta}(X)$)
can be obtained by numerically summing more components of the sum, and using
the integral bounds only on partial sums.  The pattern entropy is bounded as follows:
\begin{theorem}
\label{theorem_integer_dist}
Let $n \rightarrow \infty$.  Then,
for $\pvec$ in \eref{eq:integer_dist},
\be
 \label{eq:entropy_dist_int}
 \frac{H_{\theta} \left ( \Psi^n \right )}{n} =
 \left \{
 \begin{array}{ll}
 \left ( 1 + o(1) \right ) \cdot
 \left \{
 \frac{\alpha \ln 2}{1-\gamma}
 \left ( \log \frac{n}{2} \right )^{1-\gamma} +
 \frac{\alpha ( 1+ \gamma)}{\gamma^2}
 \left [
 \frac{1+ \gamma \ln \log 3}{(\log 3)^{\gamma}} -
 \frac{1 + \gamma \ln \log n}{(\log n)^{\gamma}} \right ] \right \}, &
 \mbox{for}~\gamma < 1, \\
 \left ( 1 + o(1) \right ) \cdot
 \left \{
 \alpha (\ln 2) \ln \log n +
 2 \alpha \left [ \frac{1 + \ln \log 3}{\log 3} - \frac{1 + \ln \log n}{\log n}
 \right ] \right \}, & \mbox{for}~\gamma = 1, \\
 H_{\theta} \left (X \right ) - \left ( 1 + o(1) \right )
 \frac{\alpha \ln 2}{(\gamma - 1) (\log n)^{\gamma -1}}, &
 \mbox{for}~\gamma > 1.
 \end{array}
 \right .
\ee
\end{theorem}
\bef
\begin{minipage}[c]{.50\textwidth}
\includegraphics
[bbllx=50pt,bblly=185pt,bburx=560pt,
bbury=598pt,height=7.5cm, width=8.5cm, clip=]{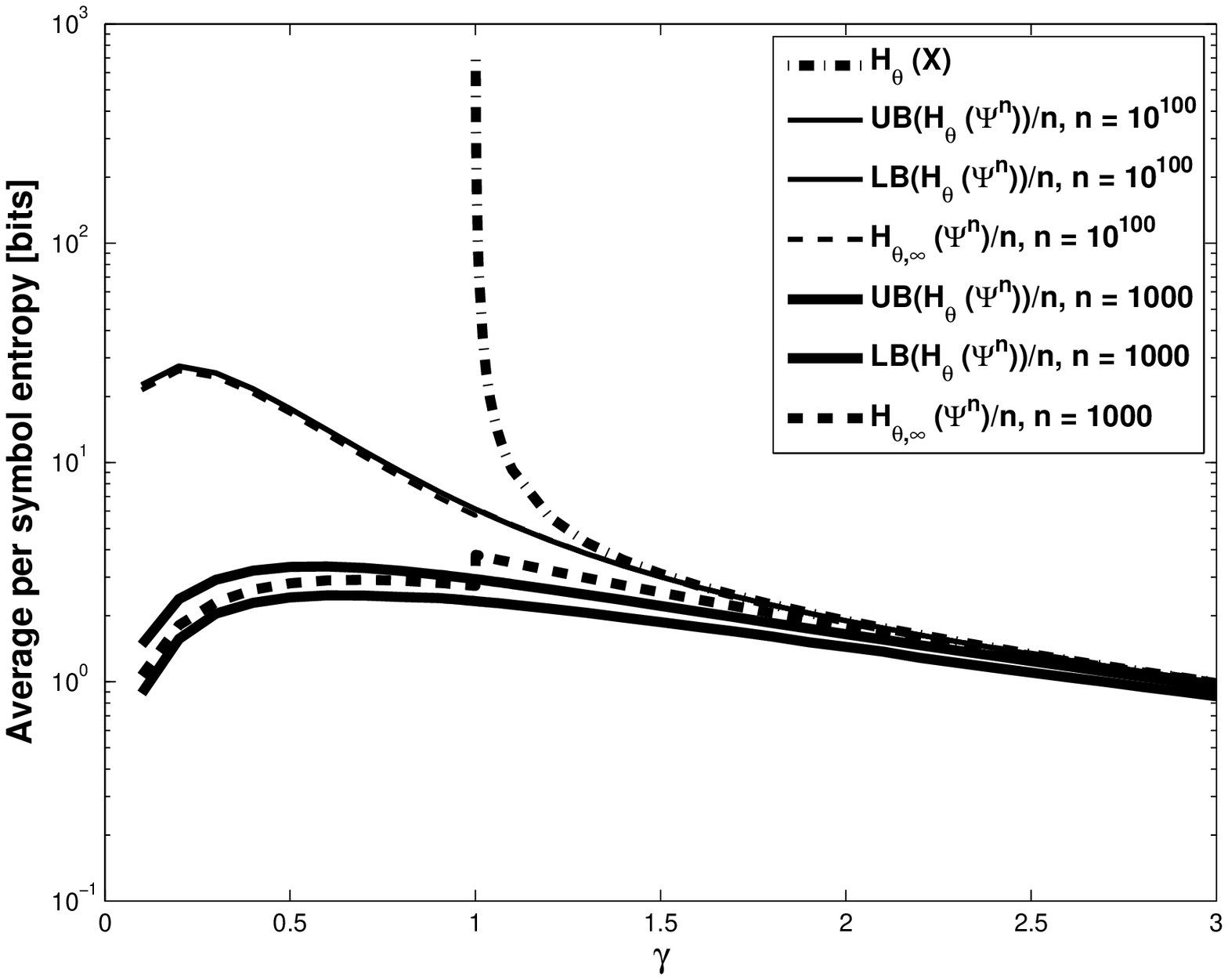}
\end{minipage}
\begin{minipage}[c]{.50\textwidth}
\includegraphics
[bbllx=50pt,bblly=185pt,bburx=560pt,
bbury=598pt,height=7.5cm, width=8.5cm, clip=]{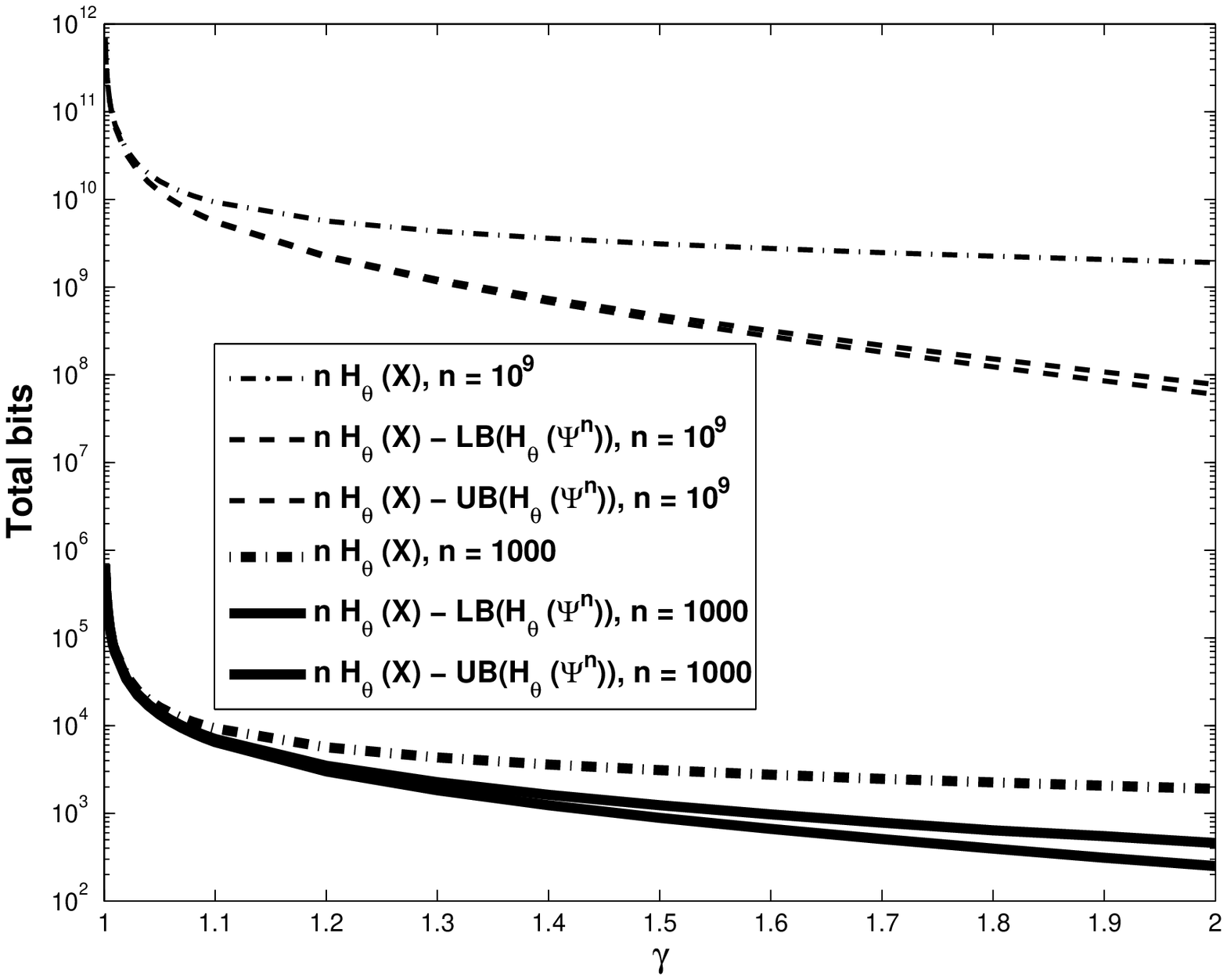}
\end{minipage}
\caption{Bounds on $H_{\theta} \left ( \Psi^n \right )$ (left) and on
$nH_{\theta}(X) - H_{\theta} \left ( \Psi^n \right )$ (right)
vs.\ $\gamma$ for different values of $n$
for the distribution in \eref{eq:integer_dist}.  Subscript $\infty$
indicates an asymptotic bound
of Theorem~\ref{theorem_integer_dist}.}
\label{fig:int}
\enf
Theorem~\ref{theorem_integer_dist} shows that the per-symbol average $H_{\theta} \left ( \Psi^n \right )/n$
is still finite even when $H_{\theta}(X) = \infty$.  Specifically, for $\gamma < 1$ it
is $\Theta \left ((\log n)^{1-\gamma} \right )$, and for $\gamma = 1$, it is $\Theta (\log \log n )$.
(For $\gamma < 1$, a looser lower bound of the same order of magnitude was independently shown in
\cite{gemelos06}.)
The bounds in \eref{eq:entropy_dist_int} for $\gamma \leq 1$
include second order terms.  For $\gamma < 1$, while asymptotically in $n$ these
terms are negligible, they are not negligible for $\gamma \rightarrow 1$ (the $1/2$ factor in the logarithm
of the first term), and for $\gamma \rightarrow 0$ (the last terms).  Additional second order terms
for $\gamma \leq 1$ that are negligible even in these cases are $-\log \alpha$, and for an upper
bound, the last two terms of \eref{eq:int_ent_upper}.  For $\gamma > 1$,
$H_{\theta} \left ( \Psi^n \right )/n$ asymptotically equals $H_{\theta}(X)$ but
decreases from $H_{\theta}(X)$ by $\Theta \left ( 1/(\log n)^{\gamma-1} \right )$.

Figure~\ref{fig:int} shows the asymptotic bounds of Theorem~\ref{theorem_integer_dist}
in \eref{eq:entropy_dist_int} as well
as non-asymptotic bounds (which are derived in the proof
of Theorem~\ref{theorem_integer_dist} in Section~\ref{sec:monotonic}) for different $\gamma$ and $n$.
Curves are shown for bounds of $H_{\theta} \left ( \Psi^n \right )/n$ (left) and
of $nH_{\theta} (X) - H_{\theta} \left ( \Psi^n \right )$ (right).
As Theorem~\ref{theorem_integer_dist} and Figure~\ref{fig:int} show,
for small $\gamma$, \eref{eq:integer_dist} decays very slowly.  This results in infinite
$H_{\theta}(X)$ for $\gamma \leq 1$, but also in a very significant decrease of
$H_{\theta}\left ( \Psi^n \right )$ from
$nH_{\theta}(X)$, where specifically $H_{\theta}\left ( \Psi^n \right )$ is finite even for $\gamma \leq 1$.
While $H_{\theta}(X)$ in this region
is dominated by small probabilities, $H_{\theta} \left ( \Psi^n \right )$ is dominated
by the larger ones.
The decrease between the two is thus dominated by the fact that
small probability symbols rarely repeat.
As $\gamma$ increases, \eref{eq:integer_dist} decays faster, the process is dominated more
by the larger probabilities, and the decrease
from $nH_{\theta}(X)$ to $H_{\theta} \left ( \Psi^n \right )$ becomes asymptotically
negligilble, yet still significant for practical $n$.

\subsubsection{The Zipf Distribution - A Fast Decaying Distribution Over the Integers}

Now, consider the \emph{Zipf\/} (or zeta)
distribution over the integers (see, e.g.\ \cite{zipf35}, \cite{zipf49}) given by
\be
 \label{eq:integer_dist1}
 \tilde{\theta}_j = \frac{1}{\zeta(1+\gamma) \cdot j^{1+\gamma}}, ~j=1,2,\ldots,
\ee
where $\gamma > 0$, and $\zeta(1+\gamma)$ is the \emph{Riemann zeta-function\/} (see, e.g., \cite{edwards74}),
given by
\be
 \label{eq:zeta}
 \zeta(s) = \sum_{n=1}^\infty \frac{1}{n^s} =
 \frac{1}{\Gamma(s)}
 \int_0^{\infty} \frac{x^{s-1}}{e^x - 1} dx
\ee
for $s > 1$, where $\Gamma(s)$ is the Gamma function.
Approximating $\sum \tilde{\theta}_j = 1$ by integrals
\be
 \label{eq:int1_alpha_bounds}
 \frac{\gamma 2^{\gamma} + 1}{\gamma 2^{\gamma}} \leq
 \zeta(1+\gamma)
 \leq \frac{\gamma 2^{\gamma+1} + \gamma + 2}{\gamma 2^{\gamma+1}}
 \leq \frac{1+\gamma}{\gamma}.
\ee
The Zipf distribution is very common in natural language and rare event modeling.  The pattern
entropy for the Zipf distribution is thus
specifically interesting in compressing patterns of a previously unobserved language.  It can also
be used for estimation of the number of letters or words in a language by applying methods
such as in \cite{orlitsky07}, but on a Zipf distribution instead of a uniform one.
Unlike the distribution given in \eref{eq:integer_dist}, for every $\gamma > 0$, the distribution
in \eref{eq:integer_dist1} has a fixed entropy rate.  Bounding sums by
integrals (separating leading terms)
\be
 \label{eq:integer_dist1_entropy}
 \log \zeta(1+\gamma) + \frac{\left ( 1+ \gamma \right )}{\zeta(1+\gamma)}
 \left [ \frac{1}{2^{1+\gamma}} + \frac{\log \left ( 3^{\gamma} e \right )}{\gamma^2 3^{\gamma}}
 \right ] \dfn \underline{H_{\theta}}(X) \leq
 H_{\theta} (X) \leq \underline{H_{\theta}}(X) + \frac{(1+\gamma)\log 3}{\zeta(1+\gamma) \cdot 3^{1+\gamma}}.
\ee
The pattern entropy is bounded as follows:
\begin{theorem}
\label{theorem_integer_dist1}
Let $n \rightarrow \infty$.  Then,
for $\pvec$ in \eref{eq:integer_dist1},
\be
 \label{eq:entropy_dis1_int_ap}
 H_{\theta} \left ( \Psi^n \right ) = n H_{\theta}(X) -
 \Theta \left (n^{\frac{1}{1+\gamma}} \log n \right ).
\ee
More precisely,
\bea
 \label{eq:entropy_dist1_int}
 \lefteqn{
 n H_{\theta} (X) -
 \left ( 1 + \frac{1}{\gamma}  - \frac{1}{3(1+2\gamma)} \right )
 \frac{1}{(1+\gamma) \cdot \zeta(1+\gamma)^{\frac{1}{1+\gamma}}}
 \left (1 + o(1) \right ) n^{\frac{1}{1+\gamma}} \log n \leq}\\
 \nonumber
 H_{\theta} \left ( \Psi^n \right ) &\leq&
 n H_{\theta} (X) -
 \left ( 1-\frac{1}{e} + \frac{1}{\gamma} - \frac{1}{2(1+2\gamma)} \right )
 \frac{1}{(1+\gamma) \cdot \zeta(1+\gamma)^{\frac{1}{1+\gamma}}}
 \left (1 - o(1) \right ) n^{\frac{1}{1+\gamma}} \log n.
\eea
\end{theorem}
\bef
\begin{minipage}[c]{.50\textwidth}
\includegraphics
[bbllx=50pt,bblly=185pt,bburx=560pt,
bbury=598pt,height=7.5cm, width=8.5cm, clip=]{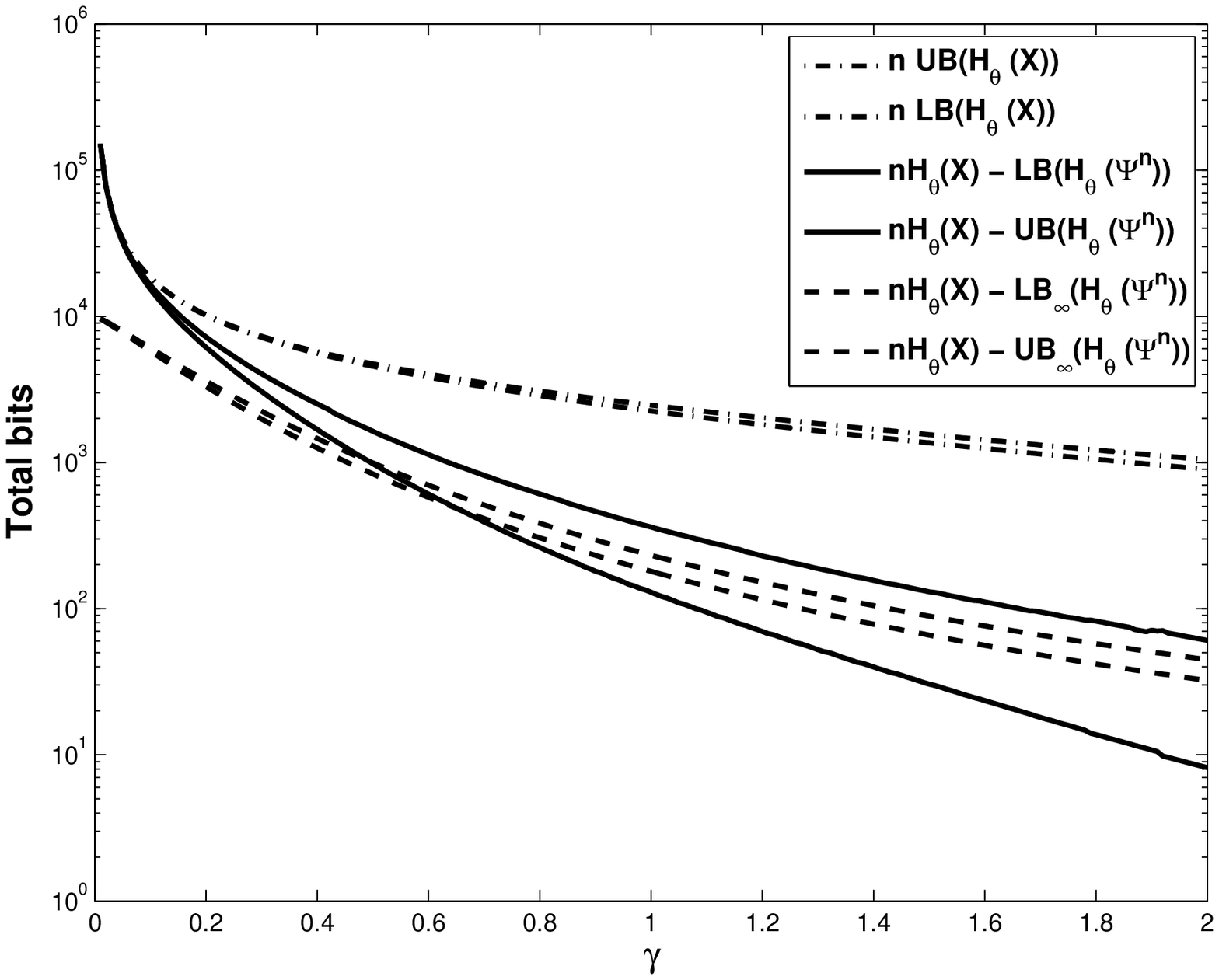}
\end{minipage}
\begin{minipage}[c]{.50\textwidth}
\includegraphics
[bbllx=50pt,bblly=185pt,bburx=560pt,
bbury=598pt,height=7.5cm, width=8.5cm, clip=]{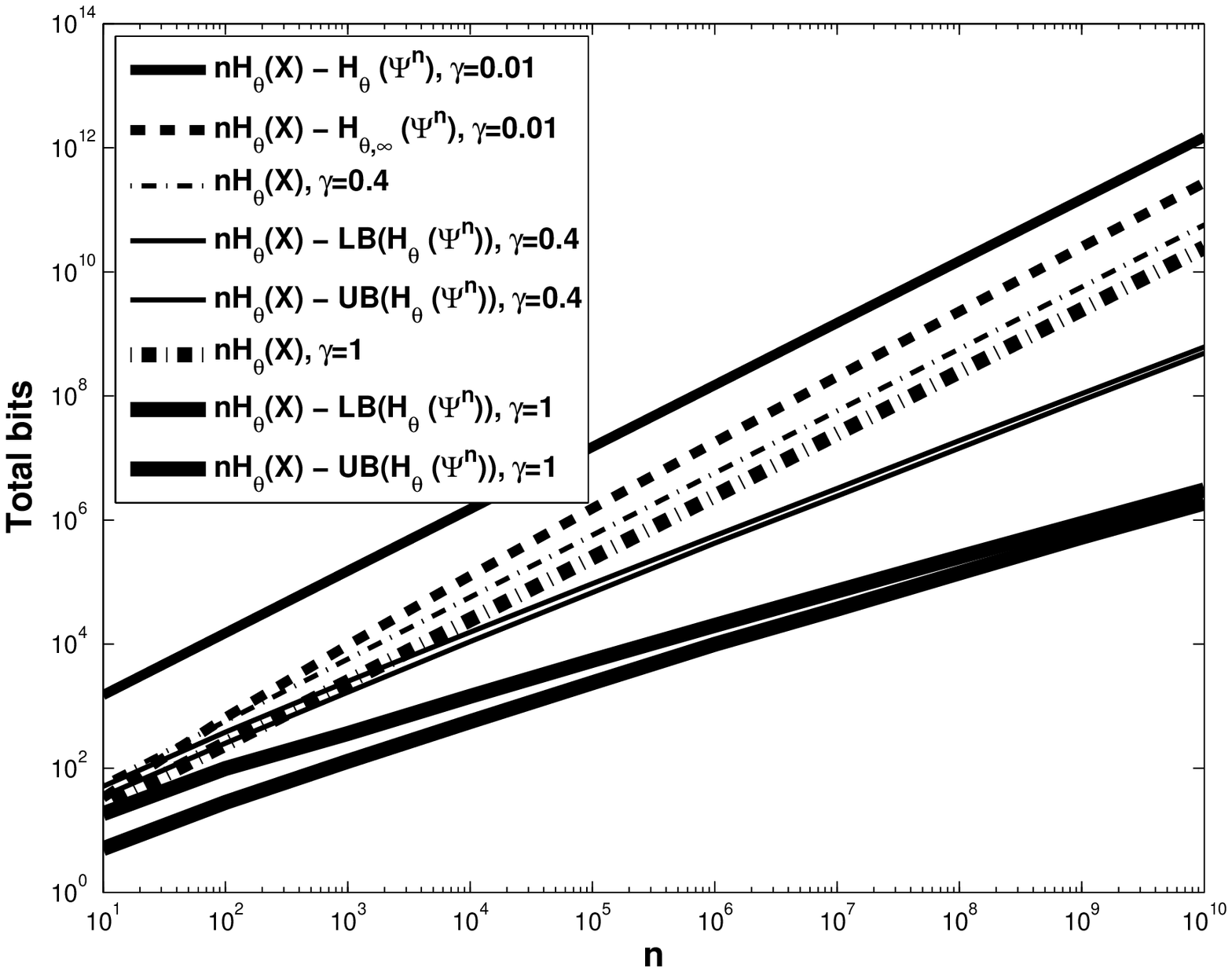}
\end{minipage}
\caption{Bounds on $nH_{\theta}(X) - H_{\theta} \left ( \Psi^n \right )$
vs.\ $\gamma$ for $n = 10^3$ (left) and vs.\ $n$ for different values of $\gamma$ (right)
for the Zipf distribution in \eref{eq:integer_dist1}.  Subscript $\infty$
indicates an asymptotic bound
of Theorem~\ref{theorem_integer_dist1}.}
\label{fig:int1}
\enf

As $\gamma$ increases, \eref{eq:integer_dist1} decays faster, and the decrease
from $nH_{\theta}(X)$ to $H_{\theta} \left ( \Psi^n \right )$ is more negligible, because
fewer letters with large enough probabilities dominate the
process.  For small $\gamma$, $H_{\theta}(X)$ is large and is
dominated mainly by symbols with relatively small probabilities.  Since such symbols rarely
repeat, $H_{\theta} \left (\Psi^n \right )$ is closer to $0$, and the
decrease from $nH_{\theta} \left (X \right )$ is thus very significant.  This behavior
resembles that of uniform distributions with $k \gg n$.
The coefficients $1$ in the lower bound and $1-1/e$ in the upper bound reflect the
effect of symbols with probabilities close to $1/n$ that may or may not occur.  The remaining
coefficients reflect decrease in entropy due to very low probability symbols, which are
unlikely to occur in $X^n$.
Figure~\ref{fig:int1} shows the asymptotic bounds of Theorem~\ref{theorem_integer_dist1}
in \eref{eq:entropy_dist1_int} as well
as non-asymptotic bounds (which are derived in the proof
of Theorem~\ref{theorem_integer_dist1} in Section~\ref{sec:monotonic}) for different $\gamma$ and $n$.
The gaps between the asymptotic and non-asymptotic behaviors are greater for smaller $\gamma$ and
smaller for greater $\gamma$. For small $n$,
second order terms are more significant.  However, for larger $n$, the gaps between the bounds
become negligible.  (Specifically, only for $\gamma = 0.01$, the asymptotic curves
do not overlap the non-asymptotic ones on the right graph.  For such low $\gamma$, curves for
lower and upper bounds do overlap.)

\subsubsection{Geometric Distribution}

The geometric distribution, which decays faster than the preceding distributions, is given by
\be
 \label{eq:geometric}
 \tilde{\theta}_j = p \left ( 1 - p \right )^{j-1};~~j = 1,2, \ldots
\ee
where $0 < p < 1$.  It has a fixed entropy rate
$H_{\theta}(X) = h_2(p)/p$, where $h_2(p)$ is the binary entropy function.  Its pattern entropy
is bounded as follows.
\begin{theorem}
\label{theorem_geometric_entropy}
Fix $p$.
Let $n \rightarrow \infty$ and let $\delta > (\ln 20)/(\ln \ln n)$.
Then, for $\pvec$ in \eref{eq:geometric},
\bea
 \nonumber
 \lefteqn{nH_{\theta} (X) - \frac{(1+\delta)^2 (\log \ln n)^2}{2\log \frac{1}{1-p}} -
  C_{L1} (p) (1+\delta) \log \ln n -
  C_{L2} (p) -
  O \left ( \frac{1}{\log n} \right )  \leq}\\
 \nonumber
 H_{\theta} \left ( \Psi^n \right ) &\leq&
 n H_{\theta} (X) -
 \left [ \frac{(1-p)h_2(p)}{p^2} - \frac{1}{2(2-p)p} \left (
 1+\frac{\log \log \log n}{\log\log n}
 \right )\right ] \cdot \frac{1}{\log \log n} + \\
 \label{eq:entropy_geometric}
 & &
 \frac{1}{2(2-p)p} \log \left ( \frac{2e(2-p)}{(1-p)\log \frac{1}{1-p}} \right )
 \cdot \frac{1}{(\log \log n)^2} +
 O \left (\frac{1}{(\log n) (\log \log n)^2} \right )
\eea
where
\bea
 \label{eq:geo_CL1}
 C_{L1}(p) &=& \frac{\log p}{\log(1-p)} +
 \frac{5 + 2p -2.5p^2}{3p(2-p)} \\
 \label{eq:geo_CL2}
 C_{L2}(p) &=&
 \frac{5 + 5p -4p^2}{3p(2-p)}\log\frac{1}{p} +
 \frac{(1-p)^2}{p^2} \left ( \frac{1}{1-p} -
 \frac{2 \left ( p^2 - 2p + 0.5 \right )}{3(2-p)^2} \right ) \log \frac{1}{1-p} + \\
 \nonumber
 & &
  \log \left [
 \left \lfloor
 \frac{2\log \frac{3}{\sqrt{1-p}}}{\log\frac{1}{1-p}}
 \right \rfloor ! \right ]+
 \sum_{b=2}^{b_{g,max}(p)} \log \left [
 \left \lfloor
 \frac{2 \log \frac{b+2}{(b-1)\sqrt{1-p}}}{\log \frac{1}{1-p}}
 \right \rfloor ! \right ] +
 \log \comb{k_{\vartheta}^-(p) + k_{\vartheta}^+(p)}{k_{\vartheta}^+(p)}
\eea
and
\be
 b_{g,max}(p) \dfn
 \frac{2 + \frac{1}{\sqrt{1-p}}}{\frac{1}{\sqrt{1-p}} - 1},~~~
 k_{\vartheta}^-(p) + k_{\vartheta}^+(p) \leq
 \left \lfloor \frac{6.9\log e}{\log\frac{1}{1-p}} + 1 \right \rfloor, ~~~
 k_{\vartheta}^+(p) \leq
 \left \lfloor \frac{1.4\log e}{\log\frac{1}{1-p}} + 1 \right \rfloor.
\ee
\end{theorem}
\bef
\begin{minipage}[c]{.50\textwidth}
\includegraphics
[bbllx=50pt,bblly=185pt,bburx=560pt,
bbury=598pt,height=7.5cm, width=8.5cm, clip=]{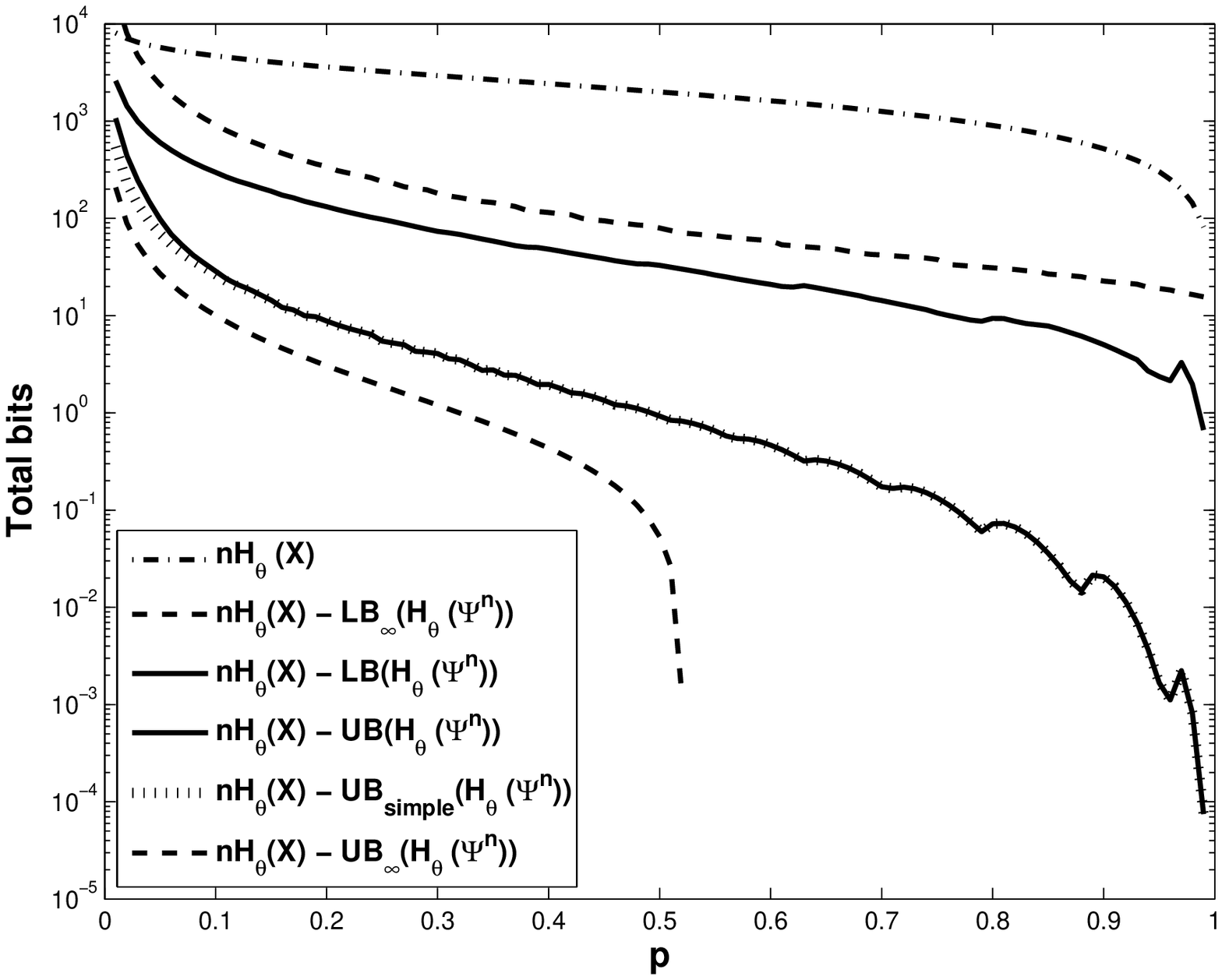}
\end{minipage}
\begin{minipage}[c]{.50\textwidth}
\includegraphics
[bbllx=50pt,bblly=185pt,bburx=560pt,
bbury=598pt,height=7.5cm, width=8.5cm, clip=]{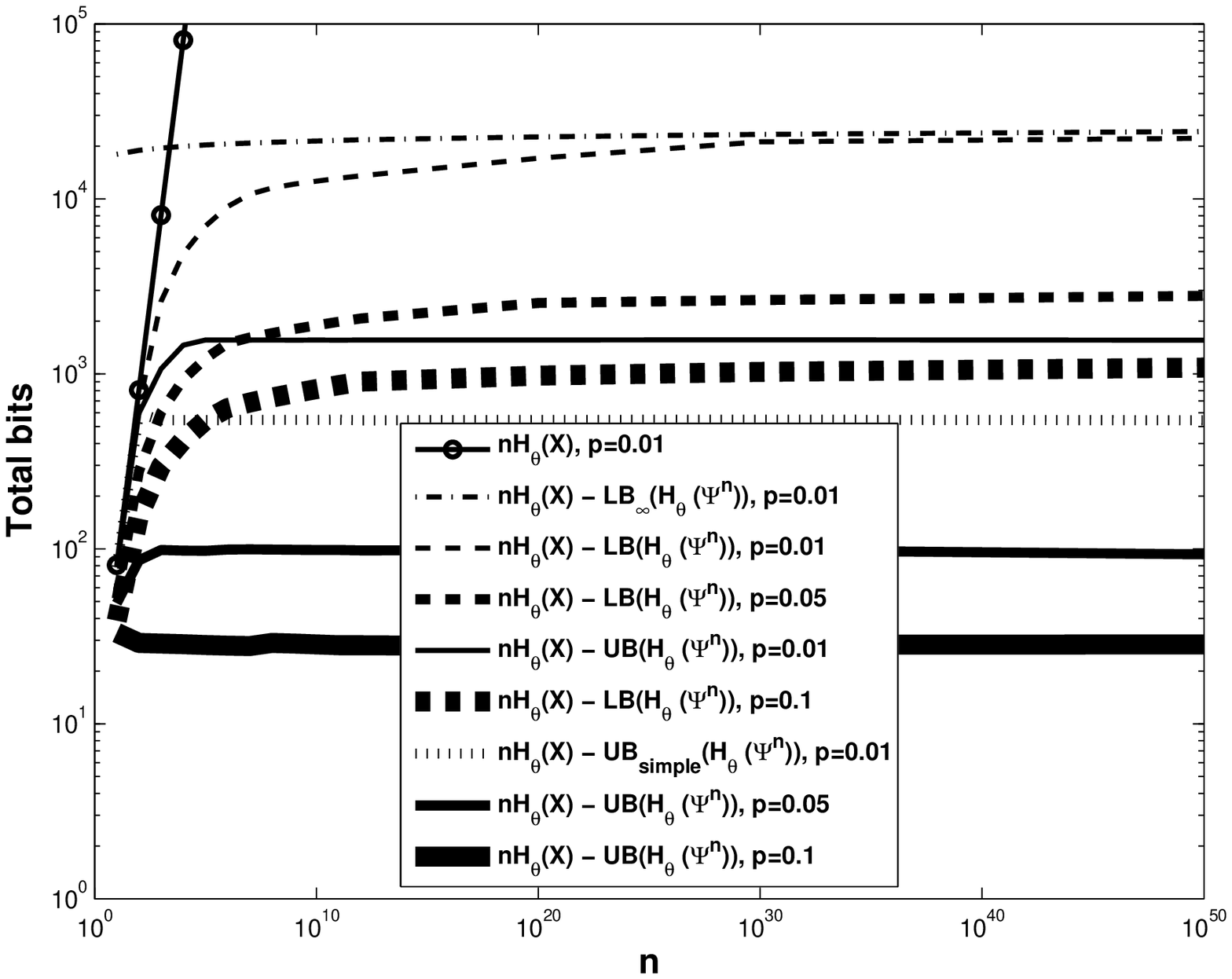}
\end{minipage}
\caption{Bounds on $nH_{\theta}(X) - H_{\theta} \left ( \Psi^n \right )$
vs.\ $p$ for $n = 10^3$ (left) and vs.\ $n$ for different values of $p$ (right)
for the geometric distribution.  Subscript $\infty$
indicates an asymptotic bound
of Theorem~\ref{theorem_geometric_entropy}, the subscript ``simple'' implies to an upper bound with
$U \geq 0$ in \eref{eq:ub3}.}
\label{fig:geo}
\enf
Theorem~\ref{theorem_geometric_entropy} shows that
$H_{\theta}\left (\Psi^n \right )$ diverges from $nH_{\theta}(X)$
by at most $\Theta \left [ (\log \log n)^2 \right ]$, and
if $p$ is smaller (for $n\rightarrow \infty$, $p \leq 0.69$),
by at least $\Theta \left ( 1/(\log \log n ) \right )$.
Due to the very slow rates, second order terms are necessary in \eref{eq:entropy_geometric}
for more accurate approximations.
The proof of Theorem~\ref{theorem_geometric_entropy}, presented
in Section~\ref{sec:monotonic}, is used to obtain numerical bounds even for relatively small $n$.
Figure~\ref{fig:geo} and Table~\ref{tab:geo_bounds}
show the asymptotic bounds of Theorem~\ref{theorem_geometric_entropy}
and the tighter non-asymptotic bounds for different $p$ and $n$.
The small bounds are very sensitive to the $\varepsilon_b$ parameters, which are
numerically chosen.  Hence, at larger $p$, where the bounds
are small, ``ringing'' appears due to quantization of $\varepsilon_b$.
A larger choice of $\delta$ above ($\delta > (\ln (20/p^2))/(\ln \ln n)$)
will eliminate the last three expressions in
\eref{eq:geo_CL2} of the asymptotic bound.  However, it will not result in
a tighter asymptotic curve.

Due to the fast decay of \eref{eq:geometric}, the decrease of $H_{\theta}\left ( \Psi^n \right )$
from $nH_{\theta}(X)$ is much smaller than in the preceding cases.  Yet, for smaller $p$,
\eref{eq:geometric} decays slower, and
$nH_{\theta}(X) - H_{\theta} \left ( \Psi^n \right )$, although negligible w.r.t.\
$nH_{\theta}(X)$ for sufficiently large $n$, is still large.
Furthermore, it is not negligible w.r.t.\
$nH_{\theta}(X)$ for smaller $n$.  Table~\ref{tab:geo_bounds} demonstrates that.
For example, for $p=0.01$, even for $n=1000$, $nH_{\theta}(X) - H_{\theta} \left ( \Psi^n \right )$
is over $10\%$ of $nH_{\theta}(X)$.  For $n=10$, $H_{\theta} \left ( \Psi^n \right ) \leq 2.28$
while $nH_{\theta}(X) > 80$.
On the other hand, for $p = 0.8$,
$nH_{\theta}(X) - H_{\theta} \left ( \Psi^n \right )$ is at most $18.66$ for $n = 10^{10}$.

As shown in Figure~\ref{fig:geo} and Table~\ref{tab:geo_bounds},
the bounds on $nH_{\theta}(X) - H_{\theta}\left (\Psi^n \right )$ are relatively
insensitive to $n$ for greater values of $n$.  This implies that the decrease
in the entropy effectively occurs during the first indices.
This is also implied by the diminishing decrease from $nH_{\theta}(X)$ on the right hand
side of \eref{eq:entropy_geometric}.
While the true rate of $nH_{\theta}(X) - H_{\theta}\left (\Psi^n \right )$
may be between those of the lower and upper bounds, diminishing decrease of
$H_{\theta} \left (\Psi^n \right )$ from $nH_{\theta}(X)$
is possible.
Fast decaying distributions may effectively behave like distributions
over small alphabets, and the gain in $H_{\theta}\left (\Psi^n \right )$ is only
due to occurrences of new indices.
Once these become sparse, we may have $H_{\theta} \left ( \Psi_{\ell} ~|~ \Psi^{\ell-1} \right ) >
H_{\theta} (X)$, thus possibly decreasing the gap between
$H_{\theta}\left (\Psi^n \right )$ and $nH_{\theta}(X)$ (as discussed in Subsection~\ref{sec:small}).

\begin{table}[htbp]
\begin{center}
\caption{Bounds on $H_{\theta} \left (\Psi^n \right )$ for different (finite) $n$.}
\label{tab:geo_bounds}
\vspace{8pt}
\begin{tabular}
{||c|c||c|c|c|c|c||}
\hline
\hline
$p$ & $n$ &
{\small $nH_{\theta}(X)$ }&
{\small UB$\left (H_{\theta}\left [ \Psi^n \right ]\right )$ }&
{\small LB$\left (H_{\theta}\left [ \Psi^n \right ]\right )$ }&
{\scriptsize
$nH_{\theta}(X)-\mbox{UB}\left (H_{\theta}\left [ \Psi^n \right ]\right )$ }&
{\scriptsize
$nH_{\theta}(X)-\mbox{LB}\left (H_{\theta}\left [ \Psi^n \right ]\right )$ } \\
\hline
\hline
$0.01$ & $10^1$  & $80.8$              & $2.28$             & $1.64$                &
         $78.52$  & $79.16$ \\
       & $10^2$  & $808$              & $212.2$              & $150.3$              &
         $595.8$ & $657.7$ \\
       & $10^3$  & $80.8 \cdot 10^2$   & $7011$ & $5483$ &
         $1069$ & $2596$ \\
       & $10^4$  & $80.8 \cdot 10^3$   & $79335$ & $75936$ &
         $1458$ & $4857$ \\
       & $10^5$  & $80.8 \cdot 10^4$   & $80.6 \cdot 10^4$ & $80.1 \cdot 10^4$ &
         $1561$ & $6979$ \\
       & $10^{10}$ & $80.8 \cdot 10^9$ & $80.8 \cdot 10^9$ & $80.8 \cdot 10^9$ &
         $1561$ & $12632$ \\
\hline
$0.05$ & $10^1$  & $57.28$ & $8.37$  & $5.31$     & $48.91$ & $51.97$ \\
       & $10^2$  & $572.8$ & $486.9$  & $295.4$ & $85.9$  & $277.4$ \\
       & $10^3$  & $5728$  & $5630$   & $5124$  & $98$    & $604$ \\
       & $10^4$  & $57280$ & $57182$  & $56348$ & $98$    & $932$ \\
\hline
$0.8$  & $10^1$  & $9.02$     & $8.96$   & $5.26$   & $0.06$ & $3.76$ \\
       & $10^2$  & $90.24$    & $90.16$  & $82.4$  & $0.08$ & $7.84$ \\
       & $10^3$  & $902.41$   & $902.34$ & $893.15$ & $0.07$ & $9.26$ \\
       & $10^{10}$& $9.02\cdot 10^9$ & $9.02\cdot 10^9$ & $9.02\cdot 10^9$ &
         $0.07$ & $18.66$ \\
\hline
\hline
\end{tabular}
\end{center}
\end{table}

\subsubsection{Linear Monotonic Distributions}

The monotonic distributions considered above were all over infinite alphabets.  Consider
a monotonic distribution over a finite alphabet, whose probabilities increase linearly.
An example of such a distribution is given by
\be
 \label{eq:lin_mono_dist}
 \theta_i = \frac{2(i-0.5)\lambda^2}{n^2}, ~i = 1,2, \ldots, k = \frac{n}{\lambda},
\ee
where $\lambda, 0 < \lambda < n$; is a parameter.  This parametrization is very similar
to that of the uniform distribution in Theorem~\ref{theorem_1n_bounds}, but here the
distribution is monotonically increasing.  For $\lambda = 1$, $k = n$, and $\theta_i < 2/n$ for all $i$.
If $\lambda \gg 1$, $k = o(n)$, and if $\lambda \ll 1$, $k \gg n$.  The i.i.d.\ entropy rate of
\eref{eq:lin_mono_dist} is
\be
 \label{eq:lin_ent}
 H_{\theta}(X) =
 \log \frac{\lambda}{n} + \log \frac{\sqrt{e}}{2} +
 O \left ( \frac{\lambda}{n} \log \frac{n}{\lambda} \right )
\ee
where the last term is negligible unless $\lambda = \Theta (n)$ (i.e., $k = \Theta(1)$).
The pattern entropy of the distribution in \eref{eq:lin_mono_dist} is as follows:
\begin{theorem}
\label{theorem_lin_entropy}
Let $n \rightarrow \infty$, let $\delta > 0$ be fixed arbitrarily small.  Then,
for $\pvec$ in \eref{eq:lin_ent}
\be
 \label{eq:entropy_lin}
 H_{\theta} \left ( \Psi^n \right ) =
 \left \{
 \begin{array}{ll}
  n H_{\theta} (X) - o(1), & \mbox{if}~ \lambda \geq n^{\frac{2}{3} + \delta} \\
  n H_{\theta} (X) - (1+o(1)) \frac{n}{\lambda} \log \frac{n}{\lambda^{3/2}}, &
  \mbox{if}~ \frac{n^{\delta}}{2} \leq \lambda \leq n^{\frac{2}{3} - \delta} \\
  (1+ o(1)) C_{\lambda} \lambda n \log \frac{n}{\lambda}, &
  \mbox{if}~\lambda \leq \frac{1}{2},
 \end{array}
 \right .
\ee
where $(1-2\lambda/3)\cdot 2/3 \leq C_{\lambda} \leq 2/3$.
\end{theorem}
Figure~\ref{fig:lin} shows the bounds for two regions of $\lambda$, and compares them
to the bounds in Corollary~\ref{cor:uniform} of a uniform distribution.  The curves include
second order terms shown in the proof of Theorem~\ref{theorem_lin_entropy}
in Section~\ref{sec:monotonic}.  Also, more complex
bounds (not shown in Section~\ref{sec:monotonic} for brevity)
obtained using Theorems~\ref{theorem:lb} and~\ref{theorem:ub3} for the boundary
between the last two regions are used.
When $k = o \left ( n^{1/3} \right )$ (first region),
there are no letters with very small probabilities.  All letters
are distributed away from each other, such that at most a single letter populates a bin.  Hence,
$H_{\theta} \left ( \Psi^n \right )$
hardly decreases from $nH_{\theta}(X)$.  When $k = o(n)$ (and is in the second region),
first occurrences of letters with
large probabilities dominate the decrease from $nH_{\theta}(X)$ to $H_{\theta} \left (\Psi^n \right )$.
The behavior is very close to that of Corollary~\ref{cor:uniform}.
However, each parameter gains $\log \left (n/\lambda^{3/2} \right )$ bits instead of $\log k = \log (n/\lambda)$
(e.g., if $\lambda = k = \sqrt{n}$, instead of $0.5 \log n$, the gain here is $0.25 \log n$).
In the last region, $H_{\theta} \left ( \Psi^n \right )= \Theta \left ( (n^2/k) \log n \right )$.
This order of magnitude, again, equals that of a uniform
distribution.
\bef
\begin{minipage}[c]{.50\textwidth}
\includegraphics
[bbllx=50pt,bblly=185pt,bburx=560pt,
bbury=598pt,height=7.5cm, width=8.5cm, clip=]{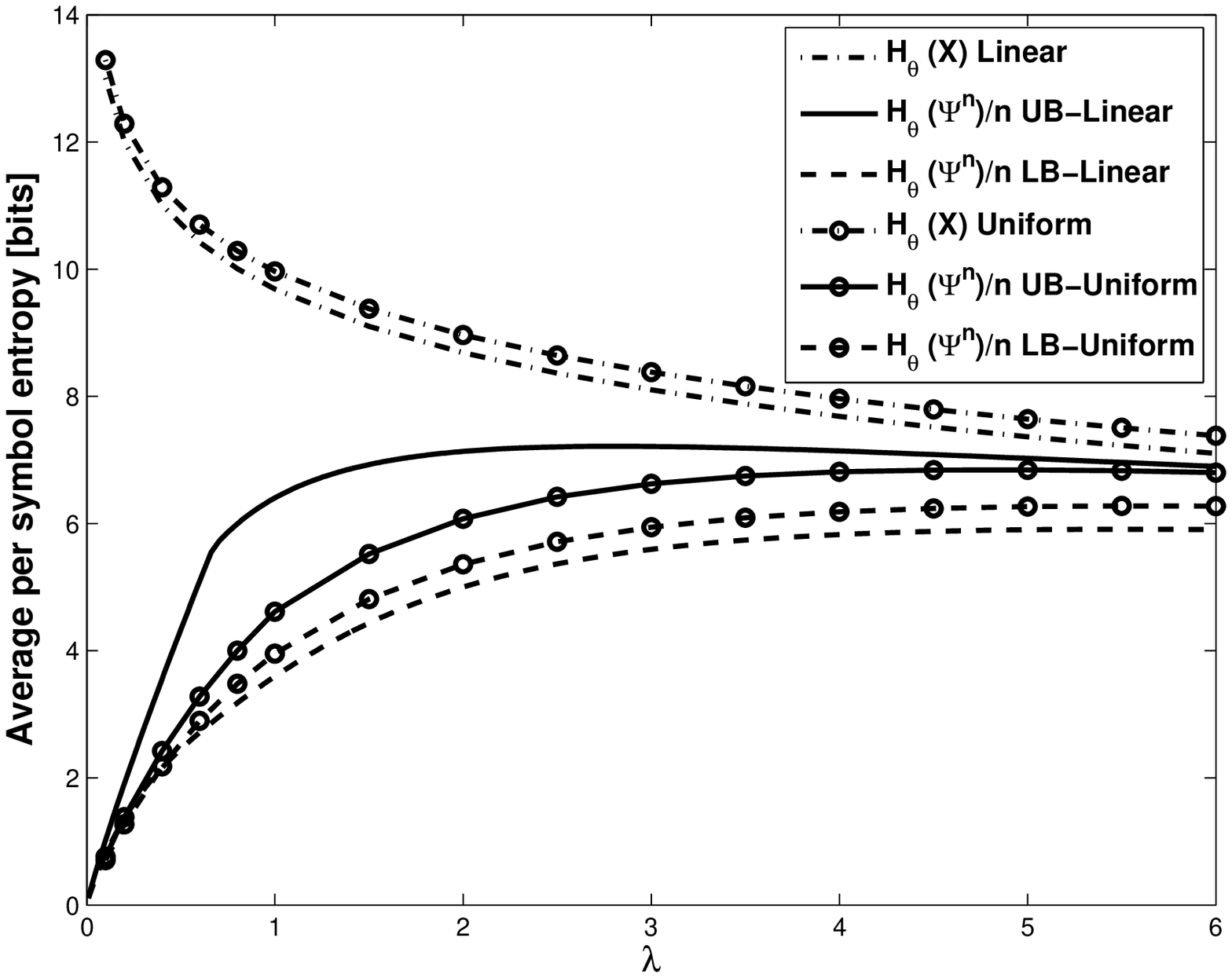}
\end{minipage}
\begin{minipage}[c]{.50\textwidth}
\includegraphics
[bbllx=50pt,bblly=185pt,bburx=560pt,
bbury=598pt,height=7.5cm, width=8.5cm, clip=]{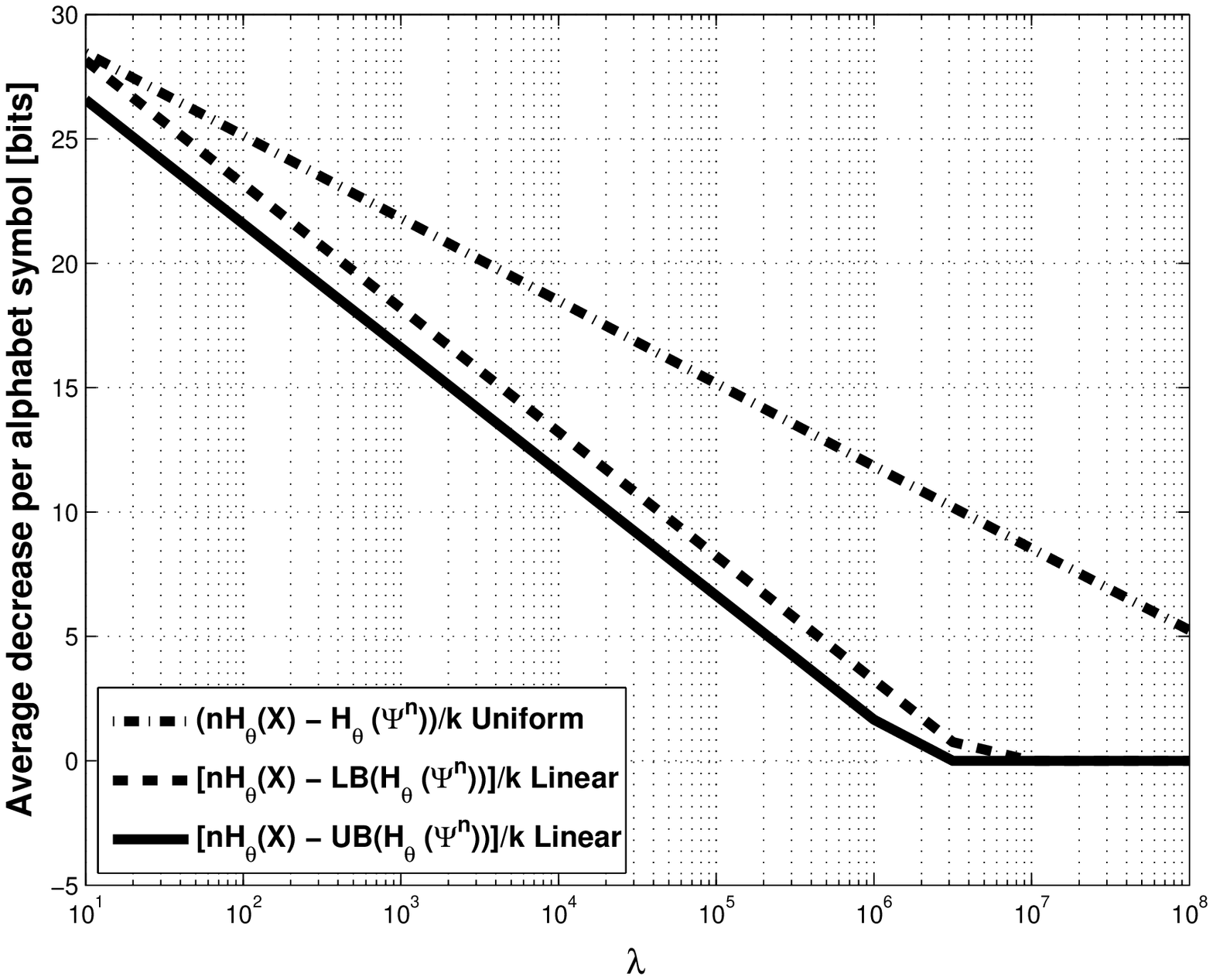}
\end{minipage}
\caption{Bounds for the linear distribution in \eref{eq:lin_mono_dist}
and the uniform distribution with $\theta_i = \lambda/n$
on $H_{\theta} \left ( \Psi^n \right )/n$
vs.\ $\lambda$
with $n = 10^3$ (left),
and on $\left [nH_{\theta}(X) - H_{\theta} \left ( \Psi^n \right ) \right ]/k$
for $n = 10^{10}$ (right).}
\label{fig:lin}
\enf

\subsection{Small Alphabets}
\label{sec:small}

While $H_{\theta} \left (\Psi^n \right ) \leq n H_{\theta}(X)$, it is not guaranteed that
$H_{\theta} \left ( \Psi_{\ell} ~|~ \Psi^{\ell-1} \right ) \leq H_{\theta}(X)$, or even
that $H_{\theta} \left ( \Psi_{n_0+1}^n ~|~ \Psi^{n_0} \right ) \leq (n - n_0) H_{\theta}(X)$
for some $n_0 < n$.  Following the chain rule
\bea
 \nonumber
 H_{\theta} \left(\Psi_{n_0+1}^n ~|~ \Psi^{n_0} \right )
 &=&
 H_{\theta} \left ( \Psi^n \right ) - H_{\theta} \left ( \Psi^{n_0} \right ) \\
 \nonumber
 &=&
 \left [ H_{\theta} \left (X^n \right ) - H_{\theta} \left ( X^n ~|~ \Psi^n \right ) \right ] -
 \left [ H_{\theta} \left (X^{n_0} \right ) - H_{\theta} \left ( X^{n_0} ~|~ \Psi^{n_0} \right ) \right ] \\
 \label{eq:pattern_cond_ent}
 &=&
 (n - n_0) H_{\theta}(X) + H_{\theta} \left ( X^{n_0} ~|~ \Psi^{n_0} \right ) -
 H_{\theta} \left ( X^n ~|~ \Psi^n \right ).
\eea
For a larger $n > n_0$, it is not guaranteed that
$H_{\theta} \left ( X^n ~|~ \Psi^n \right ) > H_{\theta} \left ( X^{n_0} ~|~ \Psi^{n_0} \right )$.
In fact, for a smaller alphabet and small $n_0$, the opposite may be true, because the longer
pattern may have less uncertainty of which symbols correspond to which indices.  This argument is in
concert with the proof of Theorem~7 in \cite{orlitsky06} and Proposition~4 in \cite{gemelos06}, which show
that for a smaller alphabet, as $n \rightarrow \infty$,
$H_{\theta} \left ( \Psi_{n+1} ~|~ \Psi^n \right ) \geq (1 - o(1))H_{\theta}(X)$.
This is true for $n \rightarrow \infty$, as long as $\theta_i > 1/n^{1-\varepsilon}$, $\forall i \leq k$;
for an arbitrarily small $\varepsilon$.

\bef
\begin{minipage}[c]{.50\textwidth}
\includegraphics
[bbllx=50pt,bblly=185pt,bburx=560pt,
bbury=598pt,height=7.5cm, width=8.5cm, clip=]{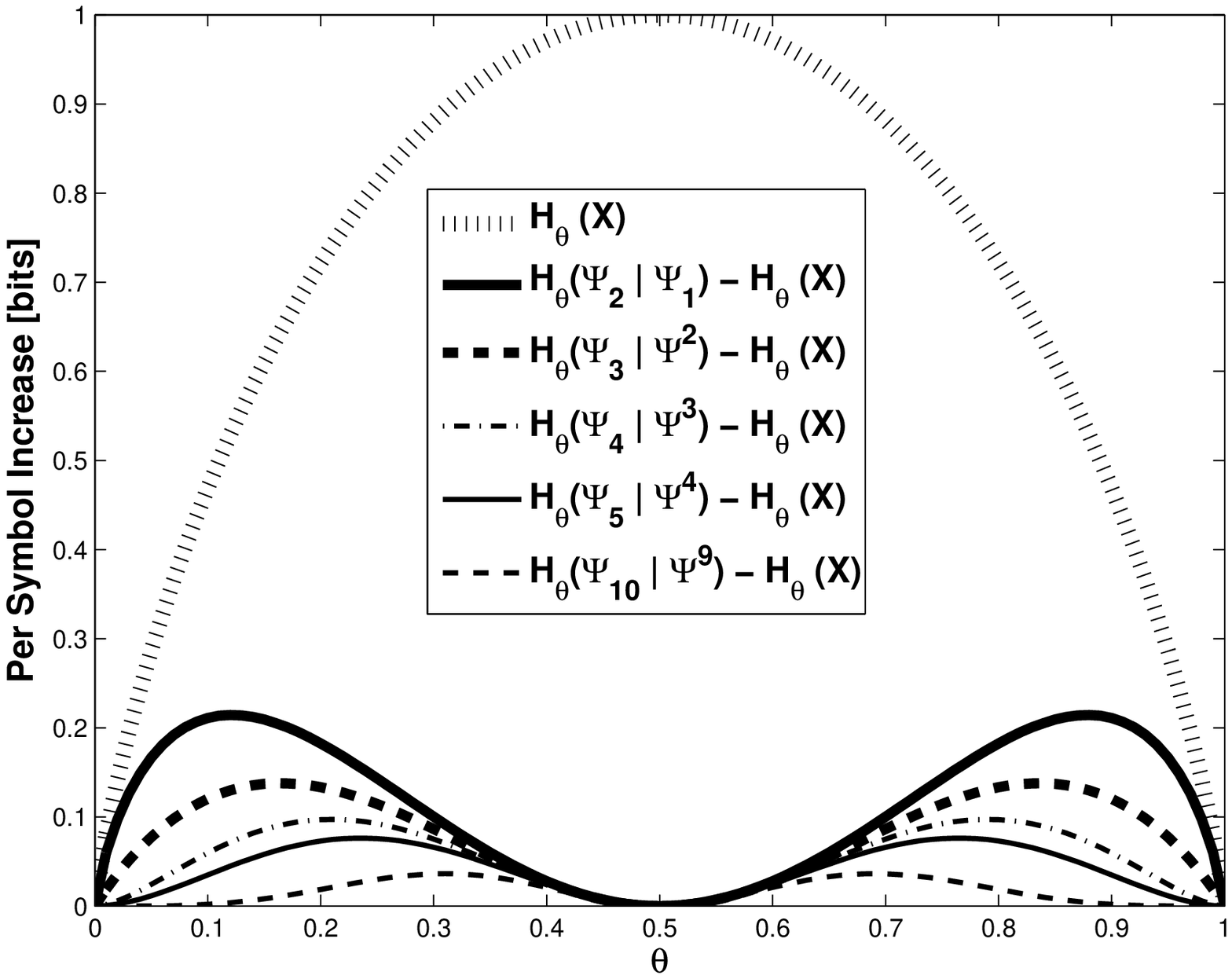}
\end{minipage}
\begin{minipage}[c]{.50\textwidth}
\includegraphics
[bbllx=50pt,bblly=185pt,bburx=560pt,
bbury=598pt,height=7.5cm, width=8.5cm, clip=]{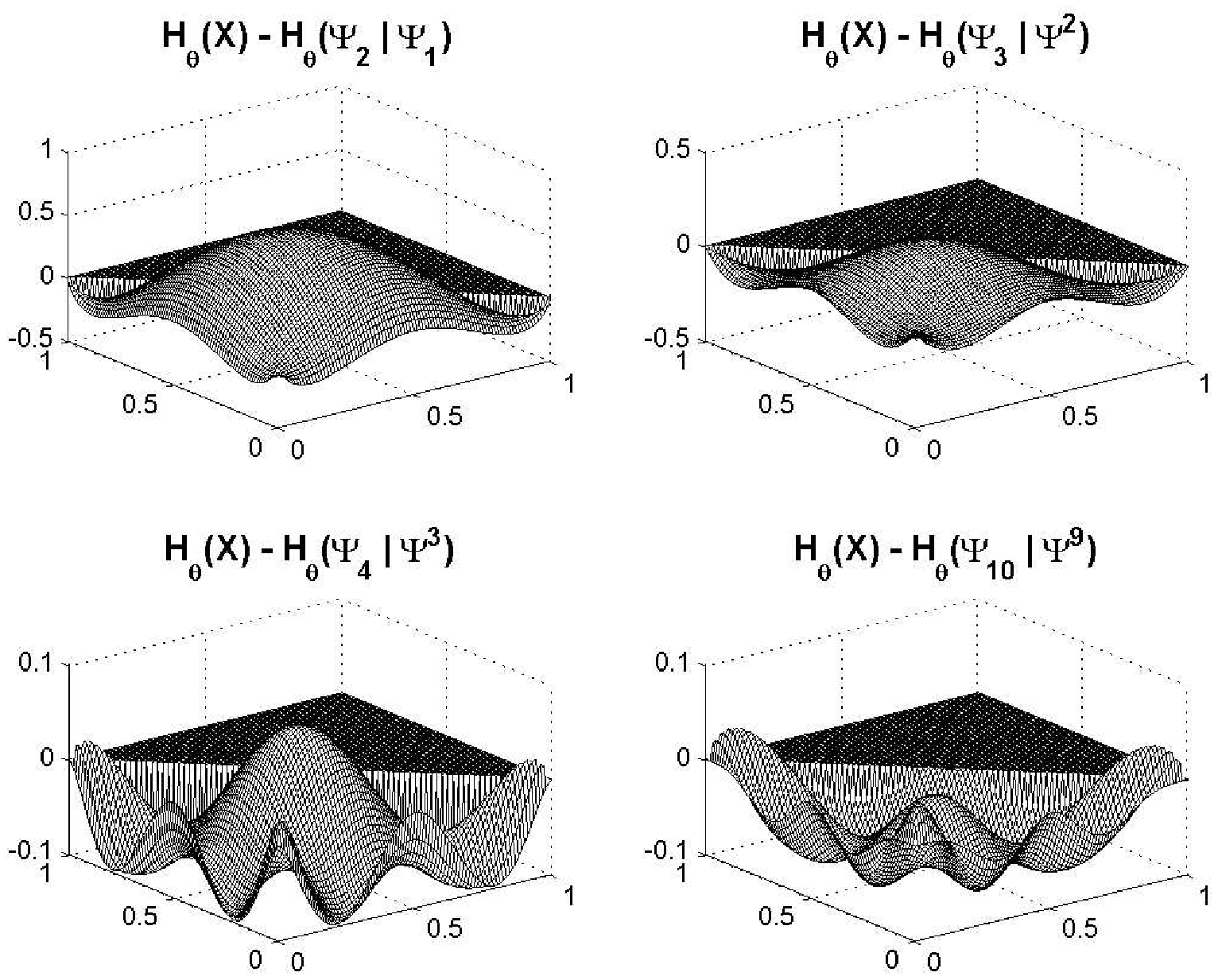}
\end{minipage}
\caption{$H_{\theta} \left ( \Psi_{\ell} ~|~ \Psi^{\ell-1} \right ) - H_{\theta}(X)$
for a binary alphabet as function of the bit probability $\theta$ (left), and
for a ternary alphabet as function of $\theta_1$ and $\theta_2$ (right).}
\label{fig:small}
\enf

Opposite behaviors, where
$H_{\theta} \left (\Psi^n \right ) \leq n H_{\theta}(X)$ but
$H_{\theta} \left ( \Psi_{\ell} ~|~ \Psi^{\ell-1} \right ) > H_{\theta}(X)$ for
$\ell > n_0$ for some $n_0 > 1$, occur for smaller alphabets because the decrease in the block entropy is
dominated by first occurrences.
Once the dominant symbols in the distribution occur, the remainder of $X^n$ consists mainly of
reoccurrences, where no decrease in entropy is exhibited.  Such a behavior can also extend
to fast decaying distributions, that while still over infinite alphabets, may only have
a small subset of the alphabet symbols that will effectively occur in a sequence, such as the
geometric distribution.  Figure~\ref{fig:small}
shows $H_{\theta} \left ( \Psi_{\ell} ~|~ \Psi^{\ell -1} \right ) - H_{\theta}(X)$ for a binary and a
ternary alphabet.  In the binary case, the decrease of $H_{\theta} \left ( \Psi^n \right )$ from
$n H_{\theta}(X)$ is the sole result of the first index, where $H_{\theta} \left (\Psi_1 \right ) = 0$.
All remaining indices have $H_{\theta} \left ( \Psi_{\ell} ~|~ \Psi^{\ell-1} \right ) \geq H_{\theta}(X)$.
Thus $n H_{\theta}(X) - H_{\theta}\left (\Psi^n \right )$ diminishes to $0$ as $n > 1$.
As shown in Figure~\ref{fig:small}, a ternary alphabet exhibits a similar behavior, except that
$H_{\theta} \left ( \Psi_{\ell} ~|~ \Psi^{\ell-1} \right ) > H_{\theta} (X)$ for the first time
at a larger $\ell$.
The value of that $\ell$ depends on the parameters of $\pvec$.
Pattern entropies shown in Figure~\ref{fig:small} were computed precisely using
\be
 \label{eq:pat_ent1}
 H_{\theta} \left ( \Psi^n \right ) = -\sum_{\nvec_x}
  \comb{n}{n_x(1),\ldots,n_x(k)} \cdot \prod_{i=1}^k \theta_i^{n_x(i)} \cdot
  \log \left \{ \sum_{\sigvec(\nvec_x)}
  \prod_{j=1}^k \theta_j^{[\sigma(\nvec_x)](j)} \right \}
\ee
where $\nvec_x \dfn \left (n_x(1), n_x(2), \ldots, n_x(k) \right )$ is the
occurrence vector of the alphabet symbols in $x^n$, the outer sum
is taken over all such vectors, and the inner sum is taken over all
$k! / \left (k - \left | \nvec_x \right | \right )!$ nonzero element
permutations $\sigvec(\nvec_x)$ of the occurrence vector, where
$\left | \nvec_x \right |$ is the cardinality of nonzero components in $\nvec_x$.
Conditional entropies were then computed with the first equality in \eref{eq:pattern_cond_ent}.

\section{Uniform Distributions - Proofs}
\label{sec:uniform}

\begin{proof}{of Corollary~\ref{cor:uniform}}
Corollary~\ref{cor:uniform} results directly from Theorems~\ref{theorem:lb}
and~\ref{theorem:ub3}.
The lower bound of \eref{eq:uniform_large_prob} is that of
\eref{eq:entropy_bounds_simple}, resulting also from
\eref{eq:lb2} and \eref{eq:lb2_S1b0}.
The upper bound follows directly from
\eref{eq:ub3} or \eref{eq:ub3_c1} with \eref{eq:ub3_U}, where $R'_b = 0$, and
the second term of \eref{eq:ub3_U} does not exist because
there is no $\theta_i$ in a bin which differs from
the average bin probability.
The lower bound of
\eref{eq:uniform_med_prob} follows from \eref{eq:lb2} with $H^{(01)}_{\theta} (X) = S_1 = S_4 = 0$.
Then, from \eref{eq:lb2_S2b1},
\bea
 \nonumber
 S_2 &\geq&
 \frac{n}{\lambda} \left ( \lambda - 1 + e^{-\lambda - \frac{\lambda^2}{n}} \right )
 \log \frac{n}{\lambda} \\
 \label{eq:uniform_med_S2}
 &\stackrel{(a)}{\geq}&
 \left ( 1 - \frac{1-e^{-\lambda}}{\lambda} \right ) n \log \frac{n}{\lambda} -
 \lambda e^{-\lambda} \log \frac{n}{\lambda}
\eea
where $(a)$ follows from $e^{-\lambda^2/n} \geq 1 - \lambda^2/n$.  Then, from \eref{eq:lb2_S3b},
\bea
 \nonumber
 S_3 &\geq&
 \left ( \log e \right ) \sum_{i=1}^{L_{01}-1} \left ( L_{01} - i \right ) \frac{\lambda}{n}
 ~=~
 \frac{\lambda \log e}{2n} \left (L_{01}^2 - L_{01} \right ) \\
  \label{eq:uniform_med_S3}
 &\stackrel{(a)}{\geq}&
 \frac{\left ( 1 - e^{-\lambda} \right )^2 \log e}{2\lambda} \cdot n -
 \frac{\left ( 1- e^{-\lambda} \right ) \log e}{2}
\eea
where $(a)$ follows from the lower bound in \eref{eq:mean_bin_bound}.
Summing \eref{eq:uniform_med_S2}-\eref{eq:uniform_med_S3} yields the lower bound
of \eref{eq:uniform_med_prob}.
The upper bound follows
from \eref{eq:ub3}, where only $R'_1$, upper bounded by \eref{eq:ub3_Rb}
using the lower bound on $L_1$ in \eref{eq:mean_bin_bound}, is
not zero.
The lower bound in \eref{eq:uniform_small_prob} follows from
\eref{eq:uniform_med_S2}-\eref{eq:uniform_med_S3} with
$\lambda = 1/n^{\mu-1+\varepsilon}$.  Expressing exponents by their Taylor series,
\bea
 \nonumber
 S_2 + S_3 &\geq&
 \left ( 1 - \frac{\lambda}{3} \right )
 \frac{\lambda n}{2} \log n^{\mu +\varepsilon} +
 \left (1 - \lambda \right ) \frac{\lambda n}{2} \log e -
 \lambda \log n^{\mu + \varepsilon} -
 \frac{\lambda \log e}{2} \\
 \nonumber
 &=&
 \left ( 1 - O \left ( \lambda + \frac{1}{n} \right ) \right )
 \frac{\lambda n }{2} \log \left (e n^{\mu +\varepsilon} \right ) \\
 &=&
 \left ( 1 - O \left ( \frac{1}{n^{\mu-1+\varepsilon}} + \frac{1}{n} \right ) \right )
 \frac{n^{2-\mu-\varepsilon}}{2} \log \left (e n^{\mu +\varepsilon}\right ).
\eea
The upper bound follows
from \eref{eq:ub3}, where only $R'_0$, which is bounded by \eref{eq:ub3_R0}, is
not zero.
\end{proof}

\begin{proof}{of Theorem~\ref{theorem_1n_bounds}} For the lower bound
\bea
 \nonumber
 H_{\theta} \left ( \Psi^n \right )
 &\stackrel{(a)}{=}&
 \left ( n - L_1 \right ) \log \frac{n}{\lambda} -
 \sum_{j=1}^{n/\lambda} P_{\theta} \left ( K_1 = j \right )
 \sum_{m=0}^{j-1} \log \left ( 1 - \frac{m\lambda}{n} \right ) \\
 \nonumber
 &\stackrel{(b)}{=}&
 n \log \frac{n}{\lambda} - \log \left [\left (\frac{n}{\lambda}\right )! \right ] +
 \sum_{j = 1}^{n/\lambda} P_{\theta} \left (K_1 = j \right )
 \log \left [ \left ( \frac{n}{\lambda} - j \right )! \right ] \\
 \nonumber
 &\stackrel{(c)}{\geq}&
 n \log \frac{n}{\lambda} - \log \left [\left (\frac{n}{\lambda}\right )! \right ] +
 \log \left [ \left (\frac{n}{\lambda} - L_1 \right )! \right ] \\
 \nonumber
 &\stackrel{(d)}{\geq}&
 n \log \frac{n}{\lambda} +
 \log \frac{\left [ \left (\frac{n}{\lambda} \right ) e^{-\lambda -\frac{\lambda^2}{n}} \right ]!}
 {\left (\frac{n}{\lambda} \right )!} \\
 &\stackrel{(e)}{=}&
 \left ( 1 - \frac{1 - e^{-\lambda}}{\lambda} \right ) n \log \frac{n}{\lambda} +
 \frac{\left (e^{\lambda} - \lambda - 1\right ) \log e}{\lambda e^{\lambda}} \cdot n -
 O \left ( \log n \right )
 \label{eq:theorem_1n_proof_lb}
\eea
Equality $(a)$ computes the average cost of repetitions (the first term) and that of
first occurrences (the second term).  Then, rearrangement of the second term leads to
$(b)$ by using $1 - m\lambda/n = (\lambda/n) \cdot (n/\lambda - m)$ and $EK_1 = L_1$.
Inequality $(c)$ is by Jensen's inequality.  Next, $(d)$ is obtained
from \eref{eq:mean_bin_bound}, and finally, Stirling's approximation
\be
 \label{eq:stirling}
 \sqrt{2 \pi m} \left ( \frac{m}{e} \right )^m \leq m! \leq
 \sqrt{2 \pi m} \left ( \frac{m}{e} \right )^m \cdot e^{1/(12m)}
\ee
and Taylor expansion of $e^{-\lambda^2/n} = 1 - O(1/n)$ are used to obtain $(e)$, proving the lower bound.

To prove the upper bound, the pattern
entropy is upper bounded by the average description length of a code that assigns
probability $\rho_1 = \lambda' / (\alpha n)$, where $\lambda' \dfn \max(\lambda, 1)$ and
$\alpha \geq 1$ is a parameter, to a repeated
index, and the remaining yet unassigned probability to a new index.  Using this code,
\bea
 \nonumber
 H_{\theta} \left ( \Psi^n \right )
 &\stackrel{(a)}{\leq}&
 \left ( n - L_1 \right ) \log \frac{\alpha n}{\lambda'} -
 \sum_{j=1}^{n/\lambda'} P_{\theta} \left ( K_1 = j \right )
 \sum_{m=0}^{j-1} \log \left ( 1 - \frac{m\lambda'}{\alpha n} \right ) \\
 \nonumber
 &\stackrel{(b)}{\leq}&
 \left ( n - L_1 \right ) \log \frac{\alpha n}{\lambda'} +
 L_1 \log \frac{\alpha n}{\lambda'} -
 \sum_{j=1}^{n/\lambda'} P_{\theta} \left ( K_1 = j \right )
 \log \frac{\left [ \frac{n}{\lambda'} \left ( \alpha - 1 \right ) + j\right ]!}
 {\left [\frac{n}{\lambda'} \left ( \alpha - 1 \right ) \right ]!} \\
 \nonumber
 &\stackrel{(c)}{\leq}&
 n \log \frac{\alpha n}{\lambda'} -
 \log \frac{\left [ \frac{n}{\lambda'} \left ( \alpha - 1 \right ) + L_1\right ]!}
 {\left [\frac{n}{\lambda'} \left ( \alpha - 1 \right ) \right ]!} \\
 \nonumber
 &\stackrel{(d)}{\leq}&
 \left ( 1 - \frac{1 - e^{-\lambda}}{\lambda} \right ) n \log \frac{n}{\lambda'} +
 \frac{\left ( 1-e^{-\lambda} \right ) \log e}{\lambda} \cdot n + \\
 \nonumber
 & &
 \left \{ \log \alpha + \frac{\alpha -1}{\lambda'} \log \left ( \alpha - 1 \right ) -
 \left ( \frac{\alpha-1}{\lambda'} + \frac{1 - e^{-\lambda}}{\lambda} \right ) \right .\\
 & &
 \label{eq:theorem_1n_proof_ub}
 \left .
 \cdot
 \log \left [ \left ( \alpha -1 \right ) + \frac{\lambda'}{\lambda} \left ( 1- e^{-\lambda} \right ) \right ]
 \right \} \cdot n +
 O \left ( \log n \right ).
\eea
Inequality $(a)$ is since the entropy is upper bounded by the average description length
of the code, which consists of the cost of repetitions (first term) and first occurrences
(second term). The bound in $(b)$ is under the worst case assumption that
all $n/\lambda'$ symbols occurred.  The first occurrence of
the last new index is assigned probability
$1 - 1/\alpha + \lambda'/(\alpha n )$,
and those of the preceding indices are assigned this probability
plus increments of $\lambda' / (\alpha n )$, depending on the occurrence time.
This step produces a tighter bound than in \eref{eq:ub3}.
Next, $(c)$ follows Jensen's inequality and the concavity of $- \log (x!)$.
Finally, $(d)$ follows Stirling's approximation and the bound
in \eref{eq:mean_bin_bound} on $L_1$.
\end{proof}

\section{Monotonic Distributions - Proofs}
\label{sec:monotonic}

\subsection{Slowly Decaying Distribution Over the Integers}

\begin{proof}{of Theorem~\ref{theorem_integer_dist}}
Let $j_0$ and $j_1$ be the indices of the greatest $\tilde{\theta}_j \leq \eta_1,~\eta_2$,
respectively.  Then, substituting $\chi_b \dfn \alpha n^{1+\varepsilon_b} (\ln 2)^{1+\gamma}$,
it can be verified that
\be
 \label{eq:int_jb}
 j_b = \left \lceil
 \frac{\beta_b \chi_b}
 {\left ( \ln \chi_b \right )^{1+\gamma}}
 \right \rceil,
 \mbox{where}~
 \beta_b = \frac{1}
 {\left [1 + \frac{\ln \beta_b}{\ln \chi_b} -
 \frac{(1+\gamma) \ln \ln \chi_b}
 {\ln \chi_b} \right ]^{1+\gamma}},~~
 b = 0, 1.
\ee
The value of $\beta_b$ can be found numerically.  It is constant for large enough $n$,
and as $n\rightarrow \infty$, it approaches $1$.  Thus
$j_b = O \left ( n^{1+\varepsilon_b}/(\log n)^{1+\gamma} \right )$.
Using an integral to approximate a sum
\be
 \label{eq:int_phi0}
 \varphi_0 = \sum_{j=j_0}^{\infty} \frac{\alpha}{j (\log j)^{1+\gamma}} =
 \frac{\alpha \ln 2}{\gamma \left ( \log j_0 \right )^{\gamma}}
 \left (1 + O \left ( \frac{1}{j_0 \log (j_0)} \right ) \right ) =
 \frac{\alpha \ln 2}{\gamma (1+\varepsilon_0)^{\gamma} \left ( \log n \right )^{\gamma}}
 \left (1 + o(1) \right ).
\ee
Similarly,
\be
 \label{eq:int_phi01}
 \varphi_{01} =
 \frac{\alpha \ln 2}{\gamma \left ( \log j_1 \right )^{\gamma}}
 \left (1 + O \left ( \frac{1}{j_1 \log (j_1)} \right ) \right ) =
 \frac{\alpha \ln 2}{\gamma (1-\varepsilon)^{\gamma} \left ( \log n \right )^{\gamma}}
 \left (1 + o(1) \right ).
\ee
Using Taylor series approximations
\be
 \label{eq:int_phi1}
 \varphi_1 = \varphi_{01} - \varphi_0 =
 \frac{\alpha (\ln 2) (\varepsilon + \varepsilon_0)}{(\log n)^{\gamma}}
 \left ( 1 + O \left ( \varepsilon + \varepsilon_0 \right ) \right )
 \left ( 1 + o(1) \right ).
\ee

To use \eref{eq:lb2}, following \eref{eq:zeroone_bin_packed_entropy}, and
selecting $\varepsilon = O( (\log \log n)/(\log n) )$,
for $\gamma < 1$,
\bea
 \nonumber
 H^{(01)}_{\theta} (X)
 &=&
 -\varphi_{01} \log \varphi_{01} - \sum_{j=2}^{j_1-1} \tilde{\theta}_j \log \tilde{\theta}_j \\
 \nonumber
 &\stackrel{(a)}{=}&
 -\varphi_{01} \log \varphi_{01} - (1 - \varphi_{01}) \log \alpha +
 \sum_{j=2}^{j_1-1} \frac{\alpha}{j(\log j)^{\gamma}} +
 \sum_{j=3}^{j_1-1} \frac{\alpha (1+\gamma) \log \log j}{j(\log j)^{1+\gamma}} \\
 \nonumber
 &\stackrel{(b)}{=}&
 - \log \alpha +
 \frac{\alpha \ln 2}{1-\gamma}
 \left ( \log \frac{j_1}{2} \right )^{1-\gamma} +
 \frac{\alpha ( 1+ \gamma)}{\gamma^2}
 \left [
 \frac{1+ \gamma \ln \log 3}{(\log 3)^{\gamma}} -
 \frac{1 + \gamma \ln \log j_1}{(\log j_1)^{\gamma}} \right ] + O(1) \\
 \label{eq:int_H01}
 &\stackrel{(c)}{=}&
 \left ( 1 + o(1) \right ) \cdot
 \left \{
 \frac{\alpha \ln 2}{1-\gamma}
 \left ( \log \frac{n}{2} \right )^{1-\gamma} +
 \frac{\alpha ( 1+ \gamma)}{\gamma^2}
 \left [
 \frac{1+ \gamma \ln \log 3}{(\log 3)^{\gamma}} -
 \frac{1 + \gamma \ln \log n}{(\log n)^{\gamma}} \right ] \right \}
\eea
where $(b)$ follows from approximating sums by integrals, substituting the value of
$\varphi_{01}$ from \eref{eq:int_phi01} and including terms equal to the last two of
the upper bound in \eref{eq:int_ent_upper} in an $O(1)$ term.  Then, $(c)$ follows from
substituting $j_1$ from \eref{eq:int_jb} with $\varepsilon = \Theta ( (\log \log n)/(\log n) )$,
and absorbing all second order terms.  Note that second order terms resulting from $j_1$ in
this step, which are absorbed in other terms, are negligible w.r.t.\ the terms expressed
above even if $\gamma \rightarrow 0$ or $\gamma \rightarrow 1$.  A similar derivation follows for
$\gamma = 1$, except that the second term in step $(b)$ is replaced by the proper
value of the integral as shown in \eref{eq:entropy_dist_int}.  In a similar manner, for
$\gamma > 1$,
\bea
 \nonumber
 H^{(01)}_{\theta} (X)
 &=&
 -\sum_{j=2}^{\infty} \tilde{\theta}_j \log \tilde{\theta}_j +
 \sum_{j=j_1}^{\infty} \tilde{\theta}_j \log \frac{\tilde{\theta}_j}{\varphi_{01}}
 ~=~
 H_{\theta}(X) -
 \sum_{j=j_1}^{\infty} \tilde{\theta}_j \log \frac{\varphi_{01}}{\tilde{\theta}_j} \\
 \nonumber
 &\stackrel{(a)}{=}&
 H_{\theta}(X) - \varphi_{01} \log \frac{\varphi_{01}}{\alpha} -
 \sum_{j=j_1}^{\infty} \frac{\alpha}{j(\log j)^{\gamma}} -
 \sum_{j=j_1}^{\infty} \frac{\alpha (1+\gamma) \log \log j}{j(\log j)^{1+\gamma}} \\
 \nonumber
 &\stackrel{(b)}{=}&
 H_{\theta}(X) - \varphi_{01} \log \frac{\varphi_{01}}{\alpha} -
 \frac{\alpha(\ln 2)}{(\gamma-1)(\log j_1)^{\gamma-1}} -
 \frac{\alpha (1+\gamma)\left ( 1 + \gamma \ln \log j_1 \right )}{\gamma^2 (\log j_1)^{\gamma}} +
 O \left ( \frac{1}{j_1 (\log j_1)^{\gamma}} \right ) \\
 \label{eq:int_H01b}
 &\stackrel{(c)}{=}&
 H_{\theta} \left (X \right ) - \left ( 1 + o(1) \right )
 \frac{\alpha \ln 2}{(\gamma - 1) (\log n)^{\gamma -1}}
\eea
where $(b)$ follows from approximating sums by integrals, and $(c)$ from substituting
$j_1$ from \eref{eq:int_jb} and absorbing second order terms, realizing that the dominant
decrease emerges from the third term.

Next, we lower bound the first sum in \eref{eq:lb2_S2b2} and $S_3$ by $0$.  Then,
choosing $\varepsilon_0 = 0$,
\be
 \label{eq:int_S2}
 S_2 \stackrel{(a)}{\geq} \sum_{j=j_1}^{j_0-1} \left ( n \tilde{\theta}_j - 1 \right )
 \log \frac{\varphi_{01}}{\tilde{\theta}_j}
 \stackrel{(b)}{\geq}
 \left ( n \varphi_1 - (j_0 - j_1) \right )
 \log \left ( \varphi_{01} n^{1-\varepsilon} \right )
 \stackrel{(c)}{=}
 O \left ( \frac{n \log \log n}{(\log n)^{\gamma}} \right )
\ee
where $(b)$ follows from $\tilde{\theta}_j \leq 1/n^{1-\varepsilon}$ in bin $1$, and
$(c)$ from \eref{eq:int_jb} and \eref{eq:int_phi1} with the choice of $\varepsilon$ and $\varepsilon_0$
above.  (Note that a tighter nontrivial bound for the second sum of \eref{eq:lb2_S2b2} can
also be obtained, but has a negligible effect.)  Next, using the trivial bound
of \eref{eq:lb2_S1b0},
\be
 \label{eq:int_S1}
 S_1 \leq \log \left ( j_1! \right ) = O \left ( \frac{n^{1-\varepsilon}}{(\log n)^{\gamma}} \right ).
\ee
Similarly, with a proper choice of constant for $\varepsilon = \Theta ((\log \log n)/(\log n))$,
$S_4 = O(j_1)$.  Adding the bounds above for all terms of \eref{eq:lb2} (normalizing $S_1$, $S_2$,
and $S_4$ by $n$) results in lower bounds satisfying \eref{eq:entropy_dist_int} for all regions
of $\gamma$, where,
regardless of $\gamma$, the expression is dominated by $H^{(01)}_{\theta}(X)$.

The bounds obtained above are asymptotic.  To derive the numerical bounds in Figure~\ref{fig:int} for
finite $n$, steps $(a)$ of \eref{eq:int_H01}, \eref{eq:int_H01b}, and \eref{eq:int_S2} are used
to compute sums (where dominant components of the sums are added, and remaining, sometimes
infinite, partial
sums are approximated by integrals).  The value of $\varepsilon$ is numerically tested for
different values, and $\varepsilon_0 = 0$ is used.  The precise expression in \eref{eq:lb2_S4} is computed
for each $\varepsilon$.  Then, $\varepsilon$ that gives
the maximal bound for each $\gamma$ and $n$ is used.  Roughly, $\varepsilon \approx 1.7 (\ln \ln n )/(\ln n)$
produced the tightest lower bounds.

Asymptotically, \eref{eq:ub3_c1} is sufficient
to obtain an upper bound on $H_{\theta} \left ( \Psi^n \right )$.
A choice of $\varepsilon = 0$ yields identical bounds to those
in \eref{eq:int_H01}-\eref{eq:int_H01b} for $H^{(01)}_\theta(X)$.
Then, the trivial bound $U \geq 0$ is used.  Finally,
\be
 \label{eq:int_R01thetasqr}
 \sum_{j = j_1}^{\infty} \tilde{\theta}_j^2 =
 \sum_{j=j_1}^{\infty} \frac{\alpha^2}{j^2 (\log j)^{2+2\gamma}}
 \stackrel{(a)}{\leq}
 \frac{\alpha^2}{(j_1 - 1) (\log j_1)^{2+2\gamma}}
 \stackrel{(b)}{=}
 O \left ( \frac{1}{n(\log n)^{1+\gamma}} \right ),
\ee
where $(a)$ follows from an integral upper bound, and $(b)$ from \eref{eq:int_jb}, yields,
using \eref{eq:ub3_R0},
\be
 \label{eq:int_R01}
 R'_{01} = O \left ( \frac{n}{(\log n)^{\gamma}} \right ).
\ee
Combining the terms of \eref{eq:ub3_c1} from \eref{eq:int_H01}-\eref{eq:int_H01b} and
\eref{eq:int_R01} yields upper bounds satisfying \eref{eq:entropy_dist_int}
dominated by $H^{(01)}_{\theta}(X)$.  The numerical upper bounds
in Figure~\ref{fig:int} can be obtained using these terms, where precise expressions
from steps $(a)$ of \eref{eq:int_H01}, \eref{eq:int_H01b} and
from \eref{eq:ub3_R0} are used to obtain
$H^{(01)}_{\theta}(X)$ and $R'_{01}$, respectively.  Slightly tighter bounds can be obtained using
\eref{eq:ub3}, where $R'_0$ and $R'_1$ are bounded separately, and $\varepsilon_0$ is numerically
optimized to minimize the bound.  (These are the bounds shown in Figure~\ref{fig:int}.)
\end{proof}

\subsection{The Zipf Distribution}

\begin{proof}{of Theorem~\ref{theorem_integer_dist1}}
For convenience, let $\alpha \dfn 1/\zeta(1 + \gamma)$.
Let $j_0$ and $j_1$ be the indices of the greatest $\tilde{\theta}_j \leq \eta_1,~\eta_2$,
respectively.
Then,
\be
 \label{eq:int1_j_limits}
 j_b = \left \lceil \alpha^{\frac{1}{1+\gamma}} \cdot
 n^{\frac{1+\varepsilon_b}{1+\gamma}} \right \rceil,~~~ b = 0,1.
\ee
Similarly,
for $b \geq 2$, define $j_b= \max \left \{1, \left \lceil
\left ( \alpha \cdot n^{1+\varepsilon_2}/ (b'+1)^2 \right )^{\frac{1}{1+\gamma}} \right \rceil
\right \}$
as the index of the greatest $\tilde{\theta}_j \leq \eta_{b+1}$
(where $b'$ is as defined in \eref{eq:eta_grid_def2}-\eref{eq:eta_grid_def}).
Note that $j_b = 1$ for $b' \geq \sqrt{\alpha n^{1+\varepsilon_2}} - 1$, $k_b = j_{b-1} - j_b$,
and some bins may be empty.
From \eref{eq:int1_j_limits} and bounding a sum by an integral,
\be
 \label{eq:int1_phi0}
 \varphi_0 = \sum_{j=j_0}^{\infty} \frac{\alpha}{j^{1+\gamma}}
 ~~~\left \{
 \begin{array}{ll}
  \leq \frac{\alpha}{j_0^{1+\gamma}} + \int_{j_0}^{\infty}
  \frac{\alpha}{x^{1+\gamma}} dx \leq
  \frac{\alpha}{\gamma j_0^{\gamma}} \left ( 1 + \frac{\gamma}{j_0} \right )\\
  \geq \int_{j_0}^{\infty}
  \frac{\alpha}{x^{1+\gamma}} dx =
  \frac{\alpha}{\gamma j_0^{\gamma}}.
 \end{array} \right .
\ee
Similarly,
\be
 \label{eq:int1_phi01}
 \frac{\alpha}{\gamma j_1^{\gamma}} \leq \varphi_{01} \leq
 \frac{\alpha}{\gamma j_1^{\gamma}} \left ( 1 + \frac{\gamma}{j_1} \right ),~~~~
 \varphi_1 = \varphi_{01} - \varphi_0 \geq \frac{\alpha}{\gamma j_1^{\gamma}}
 \left [ 1 - \left (\frac{j_1}{j_0} \right )^{\gamma} \left ( 1+ \frac{\gamma}{j_0} \right ) \right ].
\ee
From \eref{eq:int1_j_limits}-\eref{eq:int1_phi01}, it follows that
\be
 \label{eq:int1_phi_limits}
 \begin{array}{rcl}
  n\varphi_0 &=& \frac{\alpha^{\frac{1}{1+\gamma}}}{\gamma} \cdot
  n^{\frac{1-\gamma \varepsilon_0}{1+\gamma}} + O \left ( \frac{1}{n^{\varepsilon_0}} \right ), \\
  n\varphi_{01} &=& \frac{\alpha^{\frac{1}{1+\gamma}}}{\gamma} \cdot
  n^{\frac{1+\gamma \varepsilon}{1+\gamma}} + O \left ( n^{\varepsilon} \right ).
 \end{array}
\ee
While $k_0, k_{01} = \infty$, it follows from \eref{eq:int1_j_limits} that
\be
 \label{eq:int1_k1}
 k_1 = j_0 - j_1 = j_0 \left [ 1 -
 O \left ( n^{-\frac{\varepsilon_0 + \varepsilon}{1+\gamma}} \right ) \right ].
\ee

The lower bound of \eref{eq:lb2} can be derived for the distribution in \eref{eq:integer_dist1}
by separately bounding its
terms.  First, $S_3 \geq 0$.  Then, $nH^{(01)}_{\theta}(X) + S_2$ is lower bounded, and
$S_1$ and $S_4$ upper bounded.
\bea
 \nonumber
 nH_{\theta}^{(01)}(X) + S_2
 &\stackrel{(a)}{\geq}&
 nH_{\theta}(X) -
 \underbrace{n \sum_{j=j_0}^{\infty} \tilde{\theta}_j \log \frac{\varphi_{01}}{\tilde{\theta}_j}}_{V_1} -
 \underbrace{\sum_{j=j_1}^{j_0-1} \log \frac{\varphi_{01}}{\tilde{\theta}_j}}_{V_2}  \\
 \label{eq:int1_H01S2}
 & &
 + \left (1 - \frac{1}{3n^{\varepsilon_0}} - \frac{2}{n} \right )
 \underbrace{\frac{n^2}{2} \sum_{j=j_0}^{\infty} \tilde{\theta}_j^2 \log
 \frac{\varphi_{01}}{\tilde{\theta}_j}}_{V_3}
\eea
where $(a)$ follows from lower bounding \eref{eq:lb2_S2b2}, the definition of $H_{\theta}^{(01)}(X)$
in \eref{eq:zeroone_bin_packed_entropy},
and from combining terms.  Note that the summand of \eref{eq:mean_bin} can be
inserted to the summand of $V_2$ above to provide a tighter numeric expression.
Now,
\bea
\label{eq:int1_V1V2}
 V_1 + V_2
 &\stackrel{(a)}{=}&
 \left ( n \varphi_0 + j_0 - j_1 \right ) \log \frac{\varphi_{01}}{\alpha} +
 \left ( 1+\gamma \right ) \sum_{j=j_1}^{j_0-1} \log j +
 \left ( 1 + \gamma \right ) \alpha n \cdot \sum_{j = j_0}^{\infty} \frac{\log j}{j^{1+\gamma}} \\
 \nonumber
 &\stackrel{(b)}{\leq}&
 \left ( n \varphi_0 + j_0 - j_1 \right ) \log \frac{\varphi_{01}}{\alpha} +
 \left ( 1+\gamma \right ) \log \frac{j_0!}{j_1!} +
 \left ( 1 + \gamma \right ) \alpha n \cdot
 \left \{
 \frac{\log j_0}{j_0^{1+\gamma}} +
 \frac{\log j_0}{\gamma j_0^{\gamma}} +
 \frac{\log e}{\gamma^2 j_0^{\gamma}}
 \right \} \\
 \nonumber
 &\stackrel{(c)}{\leq}&
 \left ( n \varphi_0 + j_0 - j_1 \right ) \log \frac{\varphi_{01} j_0^{1+\gamma}}{\alpha} +
 \left (1+\gamma \right )
 \left [\frac{n\varphi_0 \log e}{\gamma} - \left ( j_0 - j_1 \right ) \log e \right ] +
 O \left ( j_1 \log \frac{j_0}{j_1} \right )
\eea
where $(a)$ follows from the definitions of $\varphi_0$ and $\tilde{\theta}_j$, and $(b)$
follows from bounding the sum in the last term by an integral.  The lower bound
on $\varphi_0$ in \eref{eq:int1_phi0} leads to $(c)$.  A choice of $\varepsilon_0 = 0$ leads to
the minimal tradeoff between $n\varphi_0$ and $j_0$ in the dominant term of \eref{eq:int1_V1V2}.
The smallest possible $\varepsilon$ will minimize the bound in \eref{eq:int1_V1V2}.  By
Theorem~\ref{theorem:lb}, this value is constrained to
$\varepsilon = \Theta \left [ \left ( \log \log n \right )/ \left ( \log n \right ) \right ]$ to guarantee
sufficient rate of $S_4$.  Using \eref{eq:int1_j_limits} and \eref{eq:int1_phi_limits} for
$j_b$ and $\varphi_b$, respectively,
this yields
\be
 \label{eq:int1_V1V2_final}
 V_1 + V_2 \leq
 \left ( 1 + o(1) \right ) \cdot
 \frac{\alpha^{\frac{1}{1+\gamma}}}{\gamma} \cdot
 n^{\frac{1}{1+\gamma}} \log n.
\ee
Bounding sum by an integral
\be
 \label{eq:int1_V3}
 V_3 \geq \frac{\alpha^2 n^2 j_0}{2(1+2\gamma)j_0^{2+2\gamma}} \cdot
 \log \frac{\varphi_{01} j_0^{1+\gamma}}{\alpha} +
 \frac{\alpha^2 n^2 (1+\gamma)}{2(1+2\gamma)^2} \cdot
 \frac{j_0}{j_0^{2+2\gamma}} \cdot \log e.
\ee
Using the substitutions above for $\varepsilon_0$ and $\varepsilon$, and plugging
\eref{eq:int1_V1V2_final} and \eref{eq:int1_V3} into \eref{eq:int1_H01S2},
\be
 \label{eq:int1_H01S2_final}
 nH_{\theta}^{(01)}(X) + S_2 \geq
 nH_{\theta}(X) -
 \left (1+ \frac{1}{\gamma} - \frac{1}{3(1+2\gamma)} \right )
 \frac{\alpha^{\frac{1}{1+\gamma}}}{1+\gamma}
 \left (1 + o(1) \right ) n^{\frac{1}{1+\gamma}} \log n.
\ee
Optimization that also includes the bound on $V_3$ in \eref{eq:int1_V3} yields
a slightly greater optimal $\varepsilon_0 > 0$ (roughly between $0.1$ and $0.2$) that produces the maximal
overall lower bound on $nH^{(01)}_{\theta} (X) + S_2$.  However, this bound, while more complex,
only negligibly gains on the one in \eref{eq:int1_H01S2_final} with $\varepsilon_0 = 0$.

Since $j_1 = k-k_{01}$, using the simple bound in \eref{eq:lb2_S1b0} on $S_1$, plugging
$\varepsilon = \Theta \left [ \left ( \log \log n \right )/ \left ( \log n \right ) \right ]$
\be
 \label{eq:int1_S1}
 S_1 \leq \log \left (j_1 ! \right ) =
 O \left ( j_1 \log j_1 \right ) =
 O \left ( n^{\frac{1-\varepsilon}{1+\gamma}} \log n \right ) =
 o \left ( n^{\frac{1}{1+\gamma}} \log n \right ).
\ee
In a similar manner, $S_4 = o \left ( n^{\frac{1}{1+\gamma}} \log n \right )$.  Combining
\eref{eq:int1_H01S2_final}, \eref{eq:int1_S1}, $S_3 \geq 0$, and the bound on $S_4$ into \eref{eq:lb2}
yields the lower bound in \eref{eq:entropy_dist1_int}.

The lower bound in \eref{eq:entropy_dist1_int} is asymptotic.  To obtain precise
curves as in Figure~\ref{fig:int1} for finite $n$, $j_0$ and $j_1$ are computed
with \eref{eq:int1_j_limits}.
Then, either \eref{eq:int1_phi0}-\eref{eq:int1_phi01} can be used to bound $\varphi_0$ and
$\varphi_{01}$, or they can be computed precisely substituting $j_0$ and $j_1$.
Step $(b)$ of \eref{eq:int1_V1V2} and \eref{eq:int1_V3} are used to provide
a bound on $nH_{\theta}^{(01)}(X) + S_2$, and more precise bounds are obtained on $S_1$ and
$S_4$.  (Alternatively $V_2$ can be computed precisely as
discussed following \eref{eq:int1_H01S2}.)  To obtain bounds on $S_1$,
let $\iota_b \dfn \max \left \{1, \left \lceil
\left ( \alpha \cdot n^{1-\varepsilon}/ (b+1)^2 \right )^{\frac{1}{1+\gamma}} \right \rceil
\right \}$, $b = 0,1,\ldots$;
be the index of the greatest $\tilde{\theta}_j$, such that
$\tilde{\theta}_j \leq \xi_{b+1}$.  Then,
$\kappa'_1 = \iota_0 - \iota_2$, and
\be
 \label{eq:int1_kappapb}
 \kappa'_b = \iota_{b-2} - \iota_{b+1} \leq
 \frac{\alpha^{\frac{1}{1+\gamma}} n^{\frac{1-\varepsilon}{1+\gamma}}}
 {(b-1)^{\frac{2}{1+\gamma}}} \cdot
 \frac{6}{(b-1)(1+\gamma)} + 1;~~b = 2, 3, \ldots .
\ee
This implies that only for
\be
 \label{eq:int1_S1bins}
 b \leq \left ( \frac{6}{1+\gamma} \right )^{\frac{1+\gamma}{3+\gamma}} \cdot
 \alpha^{\frac{1}{3+\gamma}} \cdot n^{\frac{1-\varepsilon}{3+\gamma}} + 1 =
 o \left ( n^{\frac{1}{3+\gamma}} \right )
\ee
there may be more than a single letter in the bins surrounding bin $b$ resulting in
nonzero summands in \eref{eq:lb2_S1b2}.  Similar derivations
can be performed to generate the elements of the sum in \eref{eq:lb2_S1b1}, and more
precise bounds on $S_4$ using \eref{eq:lb2_S4}.  Bounds are obtained for different values of $\varepsilon$,
and the value that attains a maximum is used for every $\gamma$ and $n$.
Note that $S_4$ trades off with $V_1+V_2$ by requiring a greater $\varepsilon$
to guarantee that $\varepsilon'_n$ in \eref{eq:epsipn_def} diminishes.
The choice of $\vartheta^+$ and $\vartheta^-$ also influences the tradeoff
(a smaller $\vartheta^+ - \vartheta^-$ decreases the dominant term of $S_4$ in \eref{eq:lb2_S4}).
Roughly, the optimal value of $\varepsilon$ leading to the curves
in Figure~\ref{fig:int1} equals $1.75$ to $2$ times $(\ln \ln n)/(\ln n)$ for large enough $n$.
The curves in Figure~\ref{fig:int1} were produces with
$\vartheta^- = e^{-1.97}$, and $\vartheta^+ = e^{0.98}$, that
lead to $f(\vartheta^-, \vartheta^+) > 0.2$.

To derive a tight upper bound, \eref{eq:ub3} is used, where $\etavec$ is built with
$\varepsilon = 0$ and $\varepsilon_0, \varepsilon_2 > 0$.  This is necessary for a tight bound on $R'_1$
and a negligible one on $R'_0$.
First,
\bea
 \nonumber
 nH_{\theta}^{(0,1)}(X)
 &=&
 nH_{\theta}(X) +
 \sum_{j=j_1}^{j_0-1} n \tilde{\theta}_j \log \frac{\tilde{\theta}_j}{\varphi_1} +
 \sum_{j=j_0}^{\infty} n \tilde{\theta}_j \log \frac{\tilde{\theta}_j}{\varphi_0} \\
 \nonumber
 &\stackrel{(a)}{=}&
 nH_{\theta}(X) -
 n \varphi_1 \log \frac{\varphi_1}{\alpha} -
 n \varphi_0 \log \frac{\varphi_0}{\alpha} -
 (1 + \gamma) \alpha n \sum_{j=j_1}^{\infty} \frac{\log j}{j^{1+\gamma}} \\
 \nonumber
 &\stackrel{(b)}{\leq}&
 nH_{\theta}(X) -
 n \varphi_1 \log \frac{\varphi_1}{\alpha} -
 n \varphi_0 \log \frac{\varphi_0}{\alpha} -
 (1+\gamma) \alpha n \left \{ \frac{\log j_1}{\gamma j_1^{\gamma}} +
 \frac{\log e}{\gamma^2 j_1^{\gamma}} \right \} \\
 \nonumber
 &\stackrel{(c)}{\leq}&
 nH_{\theta}(X) -
 \left ( 1 - o(1) \right ) n \varphi_1 \log \left (n \varphi_1 e^{\frac{1+\gamma}{\gamma}} \right ) -
 n \varphi_0 \log \left ( n \varphi_0 e^{\frac{1+\gamma}{\gamma}} \right ) \\
 &\stackrel{(d)}{\leq}&
 nH_{\theta}(X) - \frac{\alpha^{\frac{1}{1+\gamma}}}{\gamma (1+\gamma)} \cdot
 n^{\frac{1}{1+\gamma}} \cdot \log \frac{\alpha n
 e^{\frac{(1+\gamma)^2}{\gamma}}}{\gamma^{1+\gamma}} +
 o \left ( n^{\frac{1}{1+\gamma}} \log n \right )
 \label{eq:int1_H0_1}
\eea
where $(a)$ follows from \eref{eq:integer_dist1} and the definition of $\alpha$, $(b)$
follows from bounding a sum by an integral, $(c)$ follows from \eref{eq:int1_phi01} and \eref{eq:int1_j_limits},
and $(d)$ follows from \eref{eq:int1_phi_limits} absorbing second order terms.

To bound $R'_1$ and $R'_0$, similarly to \eref{eq:int1_phi0}-\eref{eq:int1_phi01},
\bea
 \label{eq:int1_sqsum_bin1}
 \sum_{j=j_1}^{j_0-1} \tilde{\theta}_j^2 &=&
 \sum_{j=j_1}^{j_0-1} \frac{\alpha^2}{j^{2+2\gamma}} \leq
 \frac{1}{n^2} + \frac{\alpha^2 j_1}{(1+2\gamma) j_1^{2+2\gamma}} \leq
 \frac{1}{n^2} + \frac{\alpha^{\frac{1}{1+\gamma}} n^{\frac{1}{1+\gamma}}}{(1+2\gamma)n^2}, \\
 \label{eq:int1_sqsum_bin0}
 \sum_{j=j_0}^{\infty} \tilde{\theta}_j^2 &\leq&
 \frac{1}{n^{2+2\varepsilon_0}} +
 \frac{\alpha^{\frac{1}{1+\gamma}} n^{\frac{1+\varepsilon_0}{1+\gamma}}}
 {(1+2\gamma)n^{2+2\varepsilon_0}}.
\eea
From \eref{eq:ub3_R0} and \eref{eq:int1_sqsum_bin1}-\eref{eq:int1_sqsum_bin0}, it
follows (using $k_1 \leq j_0$ and \eref{eq:int1_phi01}) that
\bea
 \label{eq:int1_R1}
 R'_1 &\leq&
 \frac{\left (1+\varepsilon_0 \right ) \alpha^{\frac{1}{1+\gamma}}\cdot n^{\frac{1}{1+\gamma}}\cdot \log n}
 {2(1+\gamma)(1+2\gamma)}  +
 \frac{\alpha^{\frac{1}{1+\gamma}}\cdot n^{\frac{1}{1+\gamma}}}{2(1+2\gamma)}  \cdot
 \log \frac{2 e (1+2\gamma) \alpha^{\frac{1}{1+\gamma}}}{\gamma} +
 O \left ( \log n \right ) \\
 \label{eq:int1_R0}
 R'_0 &=&
 O \left ( \frac{n^{\frac{1+\varepsilon_0}{1+\gamma}} \log n}{n^{2\varepsilon_0}} \right ).
\eea
While $R'_0$ requires a greater $\varepsilon_0$ to minimize its
contribution to the bound, $R'_1$ requires a smaller $\varepsilon_0$ (which implies that $k_1$
is smaller).  Trading off, a choice of
$\varepsilon_0 = \Theta [(\log \log n)/(\log n)]$ is optimal.

Finally, for bin $b \geq 2$ of $\etavec$,
\be
 \label{eq:int1_kb}
 k_b = j_{b-1} - j_b = \left ( 1 + o(1) \right )
 \alpha^{\frac{1}{1+\gamma}} n^{\frac{1+\varepsilon_2}{1+\gamma}}
 \left ( \frac{1}{b'^{\frac{2}{1+\gamma}}_b} - \frac{1}{(b'_b+1)^{\frac{2}{1+\gamma}}} \right )
\ee
where $b'_b$ is the index in $\etavec'$ as defined preceding \eref{eq:eta_grid_def}.  Specifically,
since $\eta_2 = 1/n$, $b'_1 = n^{\varepsilon_2/2} (1+o(1))$.  Following \eref{eq:mean_bin_bound} and
$\tilde{\theta}_j > \eta_2$, we have $L_b \geq (1-1/e) k_b$.  Using \eref{eq:ub3_U} and
$\varepsilon_2 = \Theta \left ( (\log \log n) / \log n \right )$,
\be
 \label{eq:int1_U1}
 U \geq (1+o(1)) \cdot \sum_{b \geq 2} L_b \log \frac{L_b}{e} \geq
 (1+o(1)) \cdot \left ( 1- \frac{1}{e} \right ) \cdot
 \sum_{b \geq 2} k_b \log \frac{(1-1/e) k_b}{e}.
\ee
Since $b'_b \geq b'_1 \rightarrow \infty$ as $n \rightarrow \infty$,
\be
 \label{eq:int1_kb1}
 k_b \geq (1+o(1)) \cdot
 \frac{\alpha^{\frac{1}{1+\gamma}} n^{\frac{1+\varepsilon_2}{1+\gamma}}}{1+\gamma}   \cdot
 \frac{1}{b'^{\frac{3+\gamma}{1+\gamma}}}.
\ee
Plugging both \eref{eq:int1_kb} and \eref{eq:int1_kb1} in \eref{eq:int1_U1},
\bea
 \nonumber
 U &\geq& (1+o(1)) \cdot \left ( 1 - \frac{1}{e} \right ) \cdot \left \{
 \sum_{b \geq 2} k_b \log \frac{(1-1/e) \alpha^{\frac{1}{1+\gamma}}
 n^{\frac{1+\varepsilon_2}{1+\gamma}}}{e(1+\gamma)} -
 \frac{(3+\gamma)\alpha^{\frac{1}{1+\gamma}}n^{\frac{1+\varepsilon_2}{1+\gamma}}}
 {(1+\gamma)^2} \cdot \sum_{b\geq 2} \frac{\log b'_b}{b'^{\frac{3+\gamma}{1+\gamma}}_b} \right \} \\
 \nonumber
 &\stackrel{(a)}{\geq}&
 (1+o(1)) \cdot \left ( 1 - \frac{1}{e} \right ) \cdot
 \frac{\alpha^{\frac{1}{1+\gamma}}n^{\frac{1}{1+\gamma}}}{1+\gamma} \log n -
 O \left ( \varepsilon_2 n^{\frac{1}{1+\gamma}} \log n \right ) \\
 \label{eq:int1_U2}
 &\stackrel{(b)}{=}&
 (1+o(1)) \cdot \left ( 1 - \frac{1}{e} \right ) \cdot
 \frac{\alpha^{\frac{1}{1+\gamma}}n^{\frac{1}{1+\gamma}}}{1+\gamma} \log n
\eea
where $(a)$ follows from \eref{eq:int1_kb} and the telescopic property of $k_b$ in $b'_b$, since
$b'^2_1 = n^{\varepsilon_2}$ and $\varepsilon_2 = \Theta \left ( (\log \log n)/\log n \right )$,
and by approximating the second sum by an integral.  Then, $(b)$ follows again from the value of $\varepsilon_2$.

Now, substituting \eref{eq:int1_H0_1},
\eref{eq:int1_R1}-\eref{eq:int1_R0}, and \eref{eq:int1_U2} in \eref{eq:ub3} yields the upper bound
in \eref{eq:entropy_dist1_int}.
Again, for the numerical bounds in Figure~\ref{fig:int1}, $\varphi_b$ and $j_b$ are computed and then
used with step $(b)$ of \eref{eq:int1_H0_1} and with \eref{eq:int1_sqsum_bin1}-\eref{eq:int1_sqsum_bin0}
and \eref{eq:ub3_R0}.  For a tighter bound on
$R'_1$, \eref{eq:ub3_Rb} can also be used directly where $L_1$ is computed
with \eref{eq:mean_bin}.  Then, $U$ is bounded with \eref{eq:ub3_U} using \eref{eq:int1_kb} to compute
$k_b$ and \eref{eq:mean_bin_bound} to compute $L_b$.
Finally, for each $\gamma$ and $n$, values of $\varepsilon_0$ and $\varepsilon_2$ that minimize the
bound are chosen.  The value of $\varepsilon_0$ is large for smaller $n$, and decreases with $n$, roughly following
the curve of $(\ln \ln n)/(\ln n)$.
This concludes the proof of
Theorem~\ref{theorem_integer_dist1}.
\end{proof}

\subsection{Geometric Distribution}
\label{sec:geometric_dist}

\begin{proof}{of Theorem~\ref{theorem_geometric_entropy}}
Let $j_0$ and $j_1$ be the indices of the greatest $\tilde{\theta}_j \leq \eta_1,~\eta_2$,
respectively.
Then,
\be
 \label{eq:geo_j_limits}
 \frac{\log \frac{pn^{1+\varepsilon_b}}{1-p}}{\log \frac{1}{1-p}}
 \leq j_b=
 \left \lceil \frac{\log \frac{pn^{1+\varepsilon_b}}{1-p}}{\log \frac{1}{1-p}} \right \rceil
 \leq
 \frac{\log \frac{pn^{1+\varepsilon_b}}{(1-p)^2}}{\log \frac{1}{1-p}},~~~ b = 0,1.
\ee
For $b \geq 2$, define $j_b= \max \left \{1, \left \lceil
\log \left \{ pn^{1+\varepsilon_2}/\left [(b'+1)^2(1-p) \right ] \right \} / \log [-(1-p)]
\right \rceil \right \}$ as the index of the greatest $\tilde{\theta}_j \leq \eta_{b+1}$
(where $b'$ is as defined in \eref{eq:eta_grid_def2}-\eref{eq:eta_grid_def}).
Note that $j_b = 1$ for $b' \geq \sqrt{pn^{1+\varepsilon_2}} - 1$, $k_b = j_{b-1} - j_b$,
and some bins may be empty.
From \eref{eq:geo_j_limits},
\be
 \label{eq:geo_phi0}
 \varphi_0 = \sum_{j=j_0}^{\infty} p (1-p)^{j-1} = (1-p)^{j_0-1}.
\ee
Similarly,
\be
 \label{eq:geo_phi01}
 \varphi_{01} = (1-p)^{j_1 - 1},~~~~
 \varphi_1 = \varphi_{01} - \varphi_0 = \varphi_{01}
 \left \{ 1 - (1-p)^{j_0 - j_1} \right \}.
\ee
From \eref{eq:geo_j_limits}-\eref{eq:geo_phi01}, it follows that
\be
 \label{eq:geo_phi_limits}
 \begin{array}{rcccl}
  \frac{1-p}{pn^{1+\varepsilon_0}} &\leq& \varphi_0 &\leq&
  \frac{1}{pn^{1+\varepsilon_0}}\\
  \frac{1-p}{pn^{1-\varepsilon}} =
  \frac{1-p}{pn^{1+\varepsilon_1}} &\leq& \varphi_{01} &\leq&
  \frac{1}{pn^{1+\varepsilon_1}} = \frac{1}{pn^{1-\varepsilon}},
 \end{array}
\ee
While $k_0, k_{01} = \infty$, if
$pn^{1+\varepsilon_1} > 1-p$, it follows from \eref{eq:geo_j_limits} that
\be
 \label{eq:geo_k1}
 \frac{\log \left [ n^{\varepsilon_0 - \varepsilon_1}(1-p)\right ]}{-\log(1-p)}
 \leq k_1 = j_0 - j_1 \leq
 \frac{\log \frac{n^{\varepsilon_0 - \varepsilon_1}}{(1-p)}}{-\log(1-p)}.
\ee
Similarly to \eref{eq:geo_phi0}-\eref{eq:geo_phi01},
\be
 \label{eq:geo_phisquare}
 \sum_{j=j_0}^{\infty} \tilde{\theta}_j^2 =
 \frac{p \varphi_0^2}{2-p},~~~~
 \sum_{j=j_1}^{\infty} \tilde{\theta}_j^2 =
 \frac{p \varphi_{01}^2}{2-p},~~~~
 \sum_{j=j_1}^{j_0-1} \tilde{\theta}_j^2 =
 \frac{p \left (\varphi_{01}^2-\varphi_0^2 \right )}{2-p}
 \stackrel{(a)}{=} \frac{p\varphi_{01}^2}{2-p}
 \left ( 1 - (1-p)^{2k_1} \right )
\ee
where $(a)$ follows from \eref{eq:geo_phi0}, \eref{eq:geo_phi01} and \eref{eq:geo_k1}.

Now, the lower bound of \eref{eq:lb2} can be derived by separately bounding its
terms.  First $S_3 \geq 0$.  Then, $nH^{(01)}_{\theta}(X) + S_2$ is lower bounded, and
$S_1$ and $S_4$ upper bounded.  Lower bounding \eref{eq:lb2_S2b2},
\bea
 \nonumber
 nH_{\theta}^{(01)}(X) + S_2
 &\geq&
 nH_{\theta}^{(01)}(X) +
 \left (1 - \frac{1}{3n^{\varepsilon_0}} - \frac{2}{n} \right )
 \underbrace{\frac{n^2}{2} \sum_{j=j_0}^{\infty} \tilde{\theta}_j^2
 \log \frac{\varphi_{01}}{\tilde{\theta}_j}}_{V_3} +
 \sum_{j = j_1}^{j_0-1} \left ( n\tilde{\theta}_j - 1 \right )
 \log \frac{\varphi_{01}}{\tilde{\theta}_j} \\
 \label{eq:geo_H01S2}
 &\stackrel{(a)}{=}&
 nH_{\theta}(X) +
 \underbrace{n \sum_{j=j_0}^{\infty} \tilde{\theta}_j \log \frac{\tilde{\theta}_j}{\varphi_{01}}}_{V_1} -
 \underbrace{\sum_{j=j_1}^{j_0-1} \log \frac{\varphi_{01}}{\tilde{\theta}_j}}_{V_2} +
 \left (1 - \frac{1}{3n^{\varepsilon_0}} - \frac{2}{n} \right )V_3
\eea
where $(a)$ follows from the definition of $H_{\theta}^{(01)}(X)$
in \eref{eq:zeroone_bin_packed_entropy} and from combining of terms.  Each component $V_{\ell}$
is now bounded.  By definition of $\tilde{\theta}_j$,
\bea
 \nonumber
 V_1 &=&
 n \varphi_0 \log \frac{p}{\varphi_{01}} +
 np(1-p) \left [ \log (1-p) \right ]
 \sum_{j = j_0}^{\infty} \left ( j - 1 \right ) \left ( 1 - p \right )^{j-2} \\
 \nonumber
 &\stackrel{(a)}{=}&
 n \varphi_0 \log \frac{p}{\varphi_{01}} +
 n \left [ \log (1-p) \right ]
 \left \{ (j_0 -1) \underbrace{(1-p)^{j_0-1}}_{\varphi_0} +
 \underbrace{(1-p)^{j_0}/p}_{(1-p)\varphi_0/p} \right \} \\
 \nonumber
 &\stackrel{(b)}{=}&
 n\varphi_0 \log \frac{\varphi_0}{\varphi_{01}} -
 \frac{n\varphi_0 h_2(p)}{p} \\
 \label{eq:geo_V1}
 &\stackrel{(c)}{\geq}&
 -\frac{\varepsilon +\varepsilon_0}{pn^{\varepsilon_0}} \log n -
 \frac{h_2(p)}{p^2 n^{\varepsilon_0}} ~=~
 - \left | O \left (\frac{(\varepsilon+\varepsilon_0) \log n}{n^{\varepsilon_0}} \right ) \right |
\eea
where $(a)$ is obtained by representing each term of the sum as a derivative of
$(1-p)^{j-1}$ w.r.t.\ $(1-p)$, exchanging order of summation and differentiation, and computing
a geometric series sum, $(b)$ follows from \eref{eq:geo_phi0}, and $(c)$ follows
from the upper bounds of \eref{eq:geo_phi_limits}
because the expression decreases with $\varphi_{01}$, and for $\varphi_0 < \varphi_{01}/e$ also
with $\varphi_0$.
From \eref{eq:geo_phi01}
\bea
 \nonumber
 V_2 &=&
 \left ( j_0 - j_1 \right ) \log \frac{(1-p)^{j_1}}{p} +
 \sum_{j=j_1}^{j_0 - 1} j \log \frac{1}{1-p} \\
 \nonumber
 &\stackrel{(a)}{=}&
 k_1 \left ( 0.5 k_1 \log \frac{1}{1-p} + \log \frac{1}{p} - 0.5\log \frac{1}{1-p} \right ) \\
 \label{eq:geo_V2}
 &\stackrel{(b)}{\leq}&
 \frac{\left ( \varepsilon_0 + \varepsilon \right )^2}{-2 \log(1-p)} (\log n)^2 +
 \left ( 0.5 + \frac{\log p}{\log(1-p)} \right ) \left (\varepsilon_0 + \varepsilon \right ) \log n +
 \log \frac{1}{p}
\eea
where $(a)$ follows from computing the sum in the second term and using the
definition of $k_1$ in \eref{eq:geo_k1}, and $(b)$ follows from the upper bound on
$k_1$ in \eref{eq:geo_k1}.
Applying similar techniques to those in \eref{eq:geo_V1},
\bea
 \nonumber
 V_3 &=&
 \frac{n^2}{2}
 \left \{
 \frac{p \varphi_0^2}{2-p} \log \frac{\varphi_{01}}{p\varphi_0} +
 \frac{(1-p)^2}{(2-p)^2} \varphi_0^2 \log \frac{1}{1-p}
 \right \} \\
 \nonumber
 &\stackrel{(a)}{\geq}&
 \frac{(1-p)^2}{2p(2-p)} \cdot
 \frac{1}{n^{2\varepsilon_0}}
 \left \{
 \left ( \varepsilon_0 + \varepsilon \right ) \log n +
 \log \frac{1}{p} +
 \left (\frac{(1-p)^2}{p(2-p)} - 1 \right ) \log \frac{1}{1-p}
 \right \} \\
 &=&
 \label{eq:geo_V3}
 O \left (\frac{(\varepsilon_0 + \varepsilon)\log n}{n^{2\varepsilon_0}} \right )
\eea
where $(a)$ follows from the lower bounds in \eref{eq:geo_phi_limits}-\eref{eq:geo_k1}.

To bound $S_1$, let $\iota_b$ be the index of the greatest $\tilde{\theta}_j$, such that
$\tilde{\theta}_j \leq \xi_{b+1}$.
Similarly to \eref{eq:geo_j_limits},
\be
 \label{eq:geo_lbj_limits}
 \iota_b = \max \left \{ 1, \left \lceil
 \frac{\log \frac{p n^{1-\varepsilon}}{(b+1)^2 (1-p)}}{\log \frac{1}{1-p}}
 \right \rceil \right \}, ~~~b = 0, 1, \ldots.
\ee
Hence, $\kappa'_1 = \iota_0 - \iota_2 \leq -2(\log 3)/\log(1-p) + 1$, and
\be
 \label{eq:geo_kappapb}
 \kappa'_b = \iota_{b-2} - \iota_{b+1} \leq
 \frac{2 \log \frac{b+2}{b-1}}{\log \frac{1}{1-p}} + 1;~~b = 2, 3, \ldots,
 \min \left ( B_{\xi}, \sqrt{pn^{1-\varepsilon}/(1-p)} - 2\right ).
\ee
For $b \geq 2$, $\kappa'_b \geq \kappa'_{b+1}$.  Hence, the maximum bound
is obtained for $b=2$, $\kappa'_2 \leq -4/\log(1-p) + 1$.  Only as long as
$\kappa'_b \geq 2$, elements of the sum in \eref{eq:lb2_S1b2} are nonzero.  This is only
possible as long as
\be
 \label{eq:geo_S1maxb}
 b \leq
 \frac{2 + \frac{1}{\sqrt{1-p}}}{\frac{1}{\sqrt{1-p}} - 1}
 \dfn b_{g,max}.
\ee
Since $k-k_{01} = j_1 -1$, from \eref{eq:epsin_def},
$\varepsilon_n \leq \min \left \{1, nj_1 e^{-0.1n^{\varepsilon}} \right \}$.
Combining these bounds, using \eref{eq:lb2_S1b2},
\bea
 \nonumber
 S_1 &\leq&
 \left (1 - \varepsilon_n \right )
 \left \{
 \log \left [
 \left \lfloor
 \frac{2\log \frac{3}{\sqrt{1-p}}}{\log\frac{1}{1-p}}
 \right \rfloor ! \right ]+
 \sum_{b=2}^{b_{g,max}} \log \left [
 \left \lfloor
 \frac{2 \log \frac{b+2}{(b-1)\sqrt{1-p}}}{\log \frac{1}{1-p}}
 \right \rfloor ! \right ]
 \right \} \\
 \label{eq:geo_S1}
 & &
 + \varepsilon_n \log (j_1!) +
 h_2 \left [ \min \left (0.5, \varepsilon_n \right ) \right ].
\eea
To guarantee that the last two terms diminish at $O \left [(\log n)^2 (\log \log n)/n \right ]$
(since $j_1 = O(\log n)$),
$\varepsilon \geq (1+\delta) (\log \ln n)/(\log n)$, where $\delta > (\ln 20)/(\ln \ln n)$ must
be used, and then, $S_1 = O(1)$.

An upper bound on $S_4$ is derived similarly to that on $S_1$.  Choosing
$\vartheta^- = e^{-5.5}$ and $\vartheta^+ = e^{1.4}$,
\bea
 \label{eq:geo_S4a}
 k_{\vartheta}^- + k_{\vartheta}^+ &\leq&
 \left \lfloor \frac{\log\frac{\vartheta^+}{\vartheta^-}}{\log \frac{1}{1-p}} + 1 \right \rfloor
 = \left \lfloor \frac{6.9\log e}{\log\frac{1}{1-p}} + 1 \right \rfloor \\
 \label{eq:geo_S4b}
 k_{\vartheta}^+ &\leq&
 \left \lfloor \frac{\log \vartheta^+}{\log \frac{1}{1-p}} + 1 \right \rfloor
 = \left \lfloor \frac{1.4\log e}{\log\frac{1}{1-p}} + 1 \right \rfloor \\
 \label{eq:geo_S4c}
 k_{\theta_i > n^{-3}} &=&
 \left \lceil \frac{\log\frac{pn^3}{1-p}}{\log \frac{1}{1-p}} \right \rceil =
 O \left ( \log n \right ).
\eea
Plugging these values in \eref{eq:lb2_S4} with the choice of $\varepsilon$ above yields
$S_4 = O(1)$, where all terms of \eref{eq:lb2_S4} but the first diminish with $n$.
(The bound can be tightened by narrowing $[\vartheta^-, \vartheta^+]$.  Such narrowing
is limited to decreasing
$f(\vartheta^-, \vartheta^+)$ in \eref{eq:epsipn_def}, such that it still produces diminishing terms
in \eref{eq:lb2_S4}.)
Note that if $\xivec$ is redefined by
$\xivec \dfn \left \{ 0, 1/n^{1-\varepsilon} \right \} \bigcup
\left \{ \tilde{\theta}_j : \tilde{\theta}_j > 1/n^{1-\varepsilon} \right \}$, and $\varepsilon$ is chosen
above with $\delta > (\ln (20/p^2))/(\ln \ln n)$, a bound of $S_1 = o(1)$ can be obtained.  This
means that for $n \rightarrow \infty$ each letter of $\pvec$ is in a single bin by itself.  A similar
approach yields $S_4 = o(1)$.  This approach, however, results in a larger
first term in an overall usually looser lower bound in \eref{eq:entropy_geometric}.

Combining \eref{eq:geo_H01S2}-\eref{eq:geo_V3}, \eref{eq:geo_S1}, and \eref{eq:lb2_S4}
gives a lower bound on $H_{\theta} \left ( \Psi^n \right )$.  Choosing $\varepsilon_0 = 0$ and
$\varepsilon = (1+\delta) (\log \ln n)/(\log n)$ with $\delta > (\ln 20)/(\ln \ln n)$
yields the lower bound
of \eref{eq:entropy_geometric}.

To numerically compute a lower bound for a finite $n$ with parameters $\varepsilon$
and $\varepsilon_0$, $j_0$ and $j_1$
are computed by \eref{eq:geo_j_limits}.
Then, \eref{eq:geo_phi0}-\eref{eq:geo_phi01} are used to compute $\varphi_0$ and $\varphi_{01}$.
Step $(b)$ of \eref{eq:geo_V1} and the first equality of \eref{eq:geo_V3}
are used to compute $V_1$ and $V_3$, respectively.
Instead of using \eref{eq:geo_V2}, the summand of \eref{eq:mean_bin}
is included in the summand of $V_2$ in \eref{eq:geo_H01S2}, and $V_2$
is precisely computed.  This is necessary for tighter bounds for very small $n$
as shown in Table~\ref{tab:geo_bounds}.
Bin count $b_{g,max}$ used in \eref{eq:geo_S1} to bound $S_1$
must be taken as the minimum
between its value in \eref{eq:geo_S1maxb} and
$\min \left (B_{\xi}, \sqrt{pn^{1-\varepsilon}/(1-p)} - 2 \right )$.
Asymptotically, the bounds of \eref{eq:lb2_S1b0} and \eref{eq:lb2_S1b1} are looser than that
of \eref{eq:lb2_S1b2} because they produce bounds of $O((\log n) (\log \log n))$ and
$O(\log n)$ on $S_1$, respectively.  However, for practical $n$, using these bounds may sometimes
produce tighter bounds.
The tightest bound for $S_1$ among those resulting from \eref{eq:lb2_S1b0}-\eref{eq:lb2_S1b2} can
be used for each $p$, $\varepsilon$, and $n$.
The sum in \eref{eq:lb2_S1b1} is bounded similarly to the sum
in \eref{eq:geo_S1}, where
the ratio $(b+2)/(b-1)$ in \eref{eq:geo_S1} is replaced by $(b+1)/b$ to bound $\kappa_b;~b=1,2,\ldots,
b_{g,max} = \min \left \{A_{\xi}, \sqrt{pn^{1-\varepsilon}/(1-p)} -1, 1/((1-p)^{-0.5} - 1) \right \}$.
Last, $S_4$ is bounded with \eref{eq:lb2_S4}, numerically computing
\eref{eq:geo_S4a}-\eref{eq:geo_S4c}.
For given $p$ and $n$, $\varepsilon$ and $\varepsilon_0$ are
numerically optimized to give the
tightest bound, resulting in the non-asymptotic curves
in Figure~\ref{fig:geo} and the values in Table~\ref{tab:geo_bounds}.
While asymptotically negligible, $S_1$ dominates the bound for
small $p$ and large $n$.
Using precise expressions instead of bounds on $V_2$ yields better bounds with larger $\varepsilon_0$.
Parameter $\varepsilon$ decreases with $n$, roughly following the curve of $1.5 (\ln \ln n)/(\ln n)$.

To derive a tight upper bound, \eref{eq:ub3} is used, where $\etavec$ is built with
$\varepsilon \leq 0$ ($\varepsilon_1 \geq 0$).  Nonnegative $\varepsilon_1$
is necessary to obtain negligible $R'_0$, yet
reducing the rate of $R'_1$.  A simpler bound can be obtained by using $U \geq 0$.  The remaining
terms of \eref{eq:ub3} are bounded below.  First,
\bea
 \nonumber
 nH_{\theta}^{(0,1)}(X)
 &=&
 nH_{\theta}(X) +
 \sum_{j=j_1}^{j_0-1} n \tilde{\theta}_j \log \frac{\tilde{\theta}_j}{\varphi_1} +
 \sum_{j=j_0}^{\infty} n \tilde{\theta}_j \log \frac{\tilde{\theta}_j}{\varphi_0} \\
 \nonumber
 &=&
 nH_{\theta}(X) +
  n \varphi_1 \log \frac{p}{\varphi_1} +
  n \varphi_0 \log \frac{p}{\varphi_0} +
  n p (1-p) \log (1-p) \sum_{j=j_1}^{\infty} (j-1) (1-p)^{j-2} \\
 \nonumber
 &\stackrel{(a)}{=}&
 nH_{\theta}(X) - \frac{n\varphi_{01}h_2(p)}{p} +
  n \varphi_1 \log \frac{\varphi_{01}}{\varphi_1} +
  n \varphi_0 \log \frac{\varphi_{01}}{\varphi_0} \\
 \nonumber
 &=&
  nH_{\theta}(X) - \frac{n\varphi_{01}h_2(p)}{p} +
  n\varphi_{01}h_2 \left (\frac{\varphi_0}{\varphi_{01}} \right ) \\
 \nonumber
 &\stackrel{(b)}{\leq}&
  nH_{\theta}(X) - \frac{(1-p) h_2(p)}{p^2 n^{\varepsilon_1}} +
  n\varphi_0 \log \frac{\varphi_{01} e}{\varphi_0} \\
 \label{eq:geo_H0_1}
 &\stackrel{(c)}{\leq}&
  nH_{\theta}(X) - \frac{(1-p) h_2(p)}{p^2 n^{\varepsilon_1}} +
  \frac{1}{pn^{\varepsilon_0}} \log
  \frac{e n^{\varepsilon_0-\varepsilon_1}}{1-p}.
\eea
where $(a)$ follows from the same reasons as $(a)$-$(b)$ in \eref{eq:geo_V1},
$(b)$ follows from \eref{eq:geo_phi_limits} and Taylor expansion on the last term,
and $(c)$ follows again from \eref{eq:geo_phi_limits}.

From \eref{eq:ub3_R0} and \eref{eq:geo_phisquare},
\bea
 \nonumber
 R'_0 &\leq&
 \frac{p}{2(2-p)} \cdot  n^2 \varphi_0^2 \cdot \log \frac{2 e(2-p)}{p\varphi_0} \\
 \label{eq:geo_R0}
 &\stackrel{(a)}{\leq}&
 \frac{1}{2(2-p)p} \cdot \frac{1}{n^{2\varepsilon_0}} \cdot
 \log \frac{2 e (2-p) n^{1+\varepsilon_0}}{1-p} ~=~
 O \left ( \frac{\log n}{n^{2\varepsilon_0}} \right )
\eea
where $(a)$ again follows from \eref{eq:geo_phi_limits}.  In a similar manner,
\bea
 \nonumber
 R'_1 &\leq&
 \frac{p}{2(2-p)} \cdot n^2 \cdot \left ( \varphi_{01}^2 - \varphi_0^2 \right ) \cdot
 \log \frac{2 e \varphi_1 k_1 (2-p)}{n p \left (\varphi_{01}^2 - \varphi_0^2 \right )} \\
 \nonumber
 &\stackrel{(a)}{\leq}&
 \frac{p\varphi_{01}^2 n^2}{2(2-p)} \cdot
 \log \frac{2 e (2-p) k_1}{np \varphi_{01}
 \left ( 1 - \frac{1}{(1-p)^2n^{2(\varepsilon_0 - \varepsilon_1)}} \right )} \\
 \nonumber
 &\stackrel{(b)}{\leq}&
 \frac{1}{2(2-p)pn^{2\varepsilon_1}} \cdot
 \left [ \varepsilon_1 \log n + \log \log n +
 \log \frac{2 e (2-p) (\varepsilon_0 - \varepsilon_1)
 \left (1 - \frac{\log(1-p)}{(\varepsilon_0-\varepsilon_1) \log n} \right )}
 {(1-p) \log \frac{1}{1-p}} \right ] \\
 \label{eq:geo_R1}
 & &
 + O \left ( \frac{1}{n^{2\varepsilon_0}} \right )
\eea
where $(a)$ follows from $\varphi_{01}^2 - \varphi_0^2 \leq \varphi_{01}^2$,
$\varphi_1 \leq \varphi_{01}$, and since $\varphi_0/\varphi_{01} = (1-p)^{j_0-j_1} \leq
1/[(1-p)n^{\varepsilon_0 - \varepsilon_1}]$ following \eref{eq:geo_phi0}-\eref{eq:geo_phi01} and
\eref{eq:geo_k1}, and $(b)$ follows from bounding $\varphi_{01}$ with
\eref{eq:geo_phi_limits} and $k_1$ with \eref{eq:geo_k1}.  Note that the logarithmic
bound on $k_1$ reduces the rate of $R'_1$.  This is the reason that two separate bins
with positive $\varepsilon_0$ and $\varepsilon_1$ are used.  With proper choices of these
parameters, $R'_0$ becomes negligible, yet, bin $0$ holds most symbols, leaving
only a logarithmic number of symbols in bin $1$.

Summing \eref{eq:geo_H0_1}, \eref{eq:geo_R0} and \eref{eq:geo_R1}, a parametric
upper bound on $H_{\theta} \left (\Psi^n \right )$ is obtained.  Substituting
a constant to $\varepsilon_0$, and letting $\varepsilon_1 = (\log \log \log n)/(\log n)$,
where $\varepsilon_0 - \varepsilon_1 \leq 1$, gives the upper bound of \eref{eq:entropy_geometric}.
The dominant terms are the first two of \eref{eq:geo_H0_1} and those of \eref{eq:geo_R1}, and
$R'_0$ is negligible.  The upper bound can be tightened
by lower bounding $U$ of \eref{eq:ub3} using \eref{eq:ub3_U}.  The
limits of the sum and its elements can be lower bounded in a similar manner to the derivation
for $S_1$ in \eref{eq:geo_lbj_limits}-\eref{eq:geo_S1}.
Since $L_b \geq k_b (1-e^{-n\tilde{\theta}_j}) = O(1/n^{\varepsilon_1}) =
O \left ( 1/(\log \log n) \right )$ when $k_b \geq 2$
this does not change the rate of the bound.
This additional term was used together with the last equality of \eref{eq:geo_H0_1} and
the first inequality of \eref{eq:geo_R0} to
produce the non-asymptotic bounds in Figure~\ref{fig:geo}, where, again,
$\varepsilon_b$ were numerically optimized.
Instead of using \eref{eq:geo_R1}, the value of $R'_1$ was computed precisely with
\eref{eq:ub3_Rb}, where $L_1$ was computed with \eref{eq:mean_bin}.  This
was necessary to achieve tight bounds for small $n$ as shown in Table~\ref{tab:geo_bounds}.
The ``simple'' bound in Figure~\ref{fig:geo} does
not include the $U$ term.  For very small $p$, this term does generate more significant gain.
For example, for $n=10^5$, and $p=0.01$, out of at least $1561$ bits of decrease from $nH_{\theta}(X)$,
$1017$ result from the term $U$ (i.e., multiple letters in bins $b > 1$ of $\etavec$).  However,
for greater $p$ the gain from $U$ diminishes, because very few bins $b>1$ (if any) contain
more than a single letter.
\end{proof}

\subsection{Linear Monotonic Distributions}

\begin{proof}{of Theorem~\ref{theorem_lin_entropy}}
Let $\varepsilon_2 = \varepsilon = \Theta \left ((\log \log n)/\log n \right ) \ll \delta /2$.  Let
$i_b$ be the smallest $i$, such that $\theta_i \geq \xi_b$, and $\ell_b$ be the smallest
$i$, such that $\theta_i \geq \eta_b$.  Hence,
\be
 \label{eq:lin_i_bin}
 i_b = \left \lceil \frac{b^2 n^{1+\varepsilon}}{2\lambda^2} + \frac{1}{2} \right \rceil,~
 b = 1,2, \ldots,~~~
 \ell_b = \left \lceil \frac{b'^2_b n^{1-\varepsilon_2}}{2\lambda^2} + \frac{1}{2} \right \rceil,~
 b = 3, 4, \ldots
\ee
where $b'_b$ is the proper index in $\etavec'$ corresponding to index $b$ in $\etavec$ (as
defined in \eref{eq:eta_grid_def}),
$\ell_2 \dfn i_1$, and
$i_0 \dfn \ell_1 = \left \lceil n^{1-\varepsilon_0}/(2\lambda^2) + 0.5 \right \rceil$.
It follows that
\be
 \label{eq:lin_phi01}
 \varphi_{01} = \left \{
 \begin{array}{ll}
 0; & \mbox{if}~\lambda \geq \sqrt{n^{1+\varepsilon}} \\
 \frac{\lambda^2}{n^2} \left ( i_1 - 1 \right )^2 =
 \frac{n^{2\varepsilon}}{4\lambda^2} \left (1 +
 O \left ( \frac{\lambda^2}{n^{1+\varepsilon}} \right ) \right ); &
 \mbox{if}~ \frac{n^{\varepsilon}}{2} \leq \lambda < \frac{\sqrt{n^{1+\varepsilon}}}{\sqrt{3}} \\
 1; &\mbox{if}~ \lambda < 0.5 n^{\varepsilon}.
 \end{array}
 \right .
\ee
In the first region, $\lambda \geq n^{2/3 + \delta}$, implying $k \leq n^{1/3 -\delta}$.
Using the trivial upper bound
$H_{\theta} \left (\Psi^n \right ) \leq nH_{\theta}(X)$.
From \eref{eq:lin_phi01}, $\varphi_{01} = 0$.  Hence, $S_2, S_3, S_4 = 0$, and
$H_{\theta}^{(01)}(X) = H_{\theta}(X)$.  Only $S_1$ remains for using \eref{eq:lb2}.  Since
$\kappa'_b = i_{b+2}-i_{b-1}$, let
\be
 \label{eq:lin_kb_r1}
 \tilde{\kappa}'_b = \frac{n^{1+\varepsilon}}{2\lambda^2}
 \left ( (b+2)^2 - (b-1)^2 \right ) =
 \frac{3n^{1+\varepsilon}}{\lambda^2}
 \left ( b + \frac{1}{2b} \right )
\ee
be the unrounded value computed to obtain $\kappa'_b$.  We must have $\tilde{\kappa}'_b \geq 1$
so that a summand in the dominant sum of \eref{eq:lb2_S1b2} is not zero.  This implies that such
summands only exist for $b \geq b'_{min}$, where
\be
 \label{eq:lin_bmin_r1}
 b'_{min} \geq \frac{\lambda^2}{3n^{1+\varepsilon}} (1+o(1))
 \stackrel{(a)}{\geq}
 \frac{1}{3}\cdot n^{\frac{1}{3} + 2\delta - \varepsilon} (1+o(1))
\ee
where $(a)$ follows from $\lambda \geq n^{2/3+\delta}$.
However, for the maximal probability, $\theta_k = 2\lambda^2(k-0.5)/n^2 \geq b^2_{max} / n^{1-\varepsilon}$.
Thus the maximal populated bin has index $b_{max} \leq \sqrt{2\lambda/n^{\varepsilon}}
\ll \lambda^2/(3n^{1 +\varepsilon})$, where the last relation follows from $\lambda \geq n^{2/3+\delta}$
and since $\varepsilon \ll \delta$.  Using \eref{eq:lb2_S1b2}, this implies that $S_1 = o(1)$.  Combining
all terms of \eref{eq:lb2}, $H_{\theta}\left ( \Psi^n \right ) \geq nH_{\theta}(X) - o(1)$.

For the second region, $S_3 \geq 0$, and lower bounding \eref{eq:lb2_S2b1},
\be
 \label{eq:lin_H01S2_r2}
 H_{\theta}^{(01)} (X) + S_2 \geq
 n H_{\theta}(X) -
 \sum_{i=1}^{i_1-1} \log \frac{\varphi_{01}}{\theta_i} =
 n H_{\theta}(X) - O \left ( i_1 \log i_1 \right ).
\ee
Similarly to \eref{eq:lin_kb_r1}, for large $b$,
\be
 \label{eq:lin_kb_r2}
 \kappa_b = i_{b+1}-i_b = (1+o(1)) \frac{n^{1+\varepsilon} b}{\lambda^2}.
\ee
Defining $\tilde{\kappa}_b$ similarly to $\tilde{\kappa}'_b$ but w.r.t.\ $\kappa_b$, and requiring
$\tilde{\kappa}_b \leq 2$ for $0$ terms in the sum of $S_1$
leads to $b_{min} \leq 2\lambda^2/n^{1+\varepsilon}$, where $b_{min}$ is
defined as $b'_{min}$ but w.r.t.\ $\kappa_b$.
Using \eref{eq:lb2_S1b1}, the sum of $S_1$ is
\bea
 \nonumber
 \sum_{b=1}^{A_{\xi}} \log \left ( \kappa_b! \right )
 &\stackrel{(a)}{=}&
 \sum_{b=b_{min}}^{b_{max}} \kappa_b \log \frac{\kappa_b}{e} +
 \frac{1}{2} \sum_{b=b_{min}}^{b_{max}} \log \kappa_b +
 O \left (b_{max} \right ) \\
 \nonumber
 &\stackrel{(b)}{=}&
 (1+o(1)) \cdot
 \left \{ \left (k - i_{b_{min}} \right ) \log \frac{n^{1+\varepsilon}}{\lambda^2 e} +
 \frac{n^{1+\varepsilon}}{\lambda^2} \sum_{b = b_{min}}^{b_{max}} b \log b \right \} \\
 \nonumber
 &\stackrel{(c)}{=}&
 (1+o(1)) \cdot
 \left \{ \left (k - i_{b_{min}} \right ) \log \frac{n^{1+\varepsilon}}{\lambda^2 e} +
 \frac{n^{1+\varepsilon}}{\lambda^2} \left ( \frac{b^2_{max}}{2} \log \frac{b_{max}}{\sqrt{e}} -
 \frac{b^2_{min}}{2} \log \frac{b_{min}}{\sqrt{e}} \right )
 \right \} \\
 \nonumber
 &\stackrel{(d)}{\leq}&
 (1+o(1)) \cdot
 \left \{
 \frac{n}{\lambda} \log \frac{n^{1+\varepsilon}}{\lambda^2 e} +
 \frac{n}{\lambda} \log \sqrt{\frac{2\lambda}{e n^{\varepsilon}}}
 \right \} \\
 \label{eq:lin_S1_r2}
 &=&
 (1+o(1)) \cdot
 \frac{n}{\lambda} \log \frac{\sqrt{2} n^{1+\varepsilon/2}}{\lambda^{3/2} e^{3/2}}
\eea
where $(a)$ follows from Stirling's approximation in \eref{eq:stirling},
$(b)$ from \eref{eq:lin_kb_r2}, $(c)$ from approximating the sum by an integral, and $(d)$ since
$k = n/\lambda$, $i_{b_{min}} \leq 2\lambda^2/n^{1+\varepsilon}$ (which follows from
\eref{eq:lin_i_bin}), and
$b_{max} \leq \sqrt{2\lambda/n^{\varepsilon}}$.  The terms that result from $i_{b_{min}}$ and
the lower limit of the integral are of second order.
(By definition of the region,
$\lambda^3 \leq n^{2-3\delta} \ll n^{2+\varepsilon}/2$, which implies that
$i_{b_{min}} \leq 2\lambda^2/n^{1+\varepsilon} \ll n/\lambda = k$.  The upper limit
on $\lambda$ also results in $b_{min} \ll b_{max}$.)

By definition in Theorem~\ref{theorem:lb} and from \eref{eq:lin_i_bin},
\be
 \label{eq:lin_S4_r2}
 S_4 = O\left (i_1 \right ) =
 O \left (\frac{n^{1+\varepsilon}}{\lambda^2} \right ) \stackrel{(a)}{=}
 o \left (\frac{n}{\lambda} \right )
\ee
where $(a)$ follows from the choice of $\varepsilon$ and since
$n^{\varepsilon} \ll n^{\delta}/2 \leq \lambda$.  Following
\eref{eq:lin_S4_r2},
the last term in \eref{eq:lin_H01S2_r2} is
$O \left (i_1 \log i_1 \right ) = o \left (S_1 \right )$.  Hence, combining all terms of
\eref{eq:lb2}
\be
 \label{eq:lin_lb_r2}
 H_{\theta} \left ( \Psi^n \right ) \geq nH_{\theta} (X) -
 (1+o(1)) \cdot
 \frac{n}{\lambda} \log \frac{3\sqrt{2} n^{1+\varepsilon/2}}{\lambda^{3/2} e^{3/2}}
\ee
where the additional $3$ in the argument of the logarithm follows from the second term
of \eref{eq:lb2_S1b1}.  With a choice of $\varepsilon = \Theta \left ( (\log \log n)/(\log n) \right )$,
this leads to the lower bound of \eref{eq:entropy_lin} in this region.

For an upper bound in the second region, $H^{(01)}_{\theta} (X) \leq H_{\theta}(X)$.  Using
\eref{eq:ub3_Rb},
\be
 \label{eq:lin_R01_r2}
 R'_{01} = O \left (n \varphi_{01} \log n \right )
 \stackrel{(a)}{=}
 O \left ( \frac{n^{1+2\varepsilon}}{\lambda^2} \log n \right )
 \stackrel{(b)}{=}
 o \left ( \frac{n}{\lambda} \right )
\ee
where $(a)$ follows from \eref{eq:lin_phi01} and $(b)$ from $n^{2\varepsilon} \ll \lambda$ in this region.
A lower bound on $U$ is obtained following the same steps as \eref{eq:lin_S1_r2}, where
$-\varepsilon_2$ replaces $\varepsilon$.  Plugging $\varepsilon_2 = \Theta \left ( (\log \log n)/(\log n) \right )$,
using \eref{eq:ub3_c1}, yields the upper bound of \eref{eq:entropy_lin}.

For the third region, let $\varepsilon_0 = -\log(2\lambda)/(\log n)$.  This leads to $\eta_1 = 2\lambda/n$.
Hence, since $\theta_k \leq 2\lambda/n$,
$H_{\theta}^{(01)} = S_1 = S_4 = U = 0$.  With looser bounding, also $S_3 \geq 0$.
From \eref{eq:lb2_S2b2}
\bea
 \nonumber
 S_2 &\geq&
 (1+o(1)) \cdot \left ( 1 - \frac{2\lambda}{3} \right ) \cdot
 \frac{n^2}{2}
 \sum_{i=1}^k \theta_i^2 \log \frac{1}{\theta_i} \\
 \nonumber
 &=&
 (1+o(1)) \cdot \left ( 1 - \frac{2\lambda}{3} \right ) \cdot
 \frac{2 \lambda^4}{n^2} \sum_{i=1}^k \left ( i-0.5 \right )^2 \log \frac{n^2}{2(i-0.5) \lambda^2} \\
 \nonumber
 &\stackrel{(a)}{=}&
 (1+o(1)) \cdot \left ( 1 - \frac{2\lambda}{3} \right ) \cdot
 \frac{2 \lambda^4 k^3}{3n^2} \log \frac{e^{1/3} n^2}{2\lambda^2 k} \\
 &\stackrel{(b)}{=}&
 (1+o(1)) \cdot \left ( 1 - \frac{2\lambda}{3} \right ) \cdot
 \frac{2}{3} \lambda n \log \frac{e^{1/3} n}{2\lambda}
 \label{eq:lin_S2_r3}
\eea
where $(a)$ follows from approximating the sum by an integral and since $n \rightarrow \infty$, and
$(b)$ from substituting $k = n/\lambda$.  To use \eref{eq:ub3_c1}, approximating a sum by an integral
\be
 \label{eq:lin_R01sq_r3}
 \sum_{i=1}^k \theta_i^2 = (1+o(1)) \cdot \frac{4\lambda^4}{n^4} \cdot \frac{k^3}{3} =
 (1+o(1)) \cdot \frac{4}{3} \cdot \frac{\lambda}{n}.
\ee
It then follows using \eref{eq:ub3_R0} that
\be
 \label{eq:lin_R01_r3}
 R'_{01} \leq (1+o(1)) \cdot \frac{2}{3} \cdot \lambda n \log \frac{3 e n}{2\lambda}.
\ee
Since all other terms but $S_2$ for the lower bound in \eref{eq:lb2}
and $R'_{01}$ for the upper bound in \eref{eq:ub3_c1} are $0$ or bounded by $0$, both
bounds are proved from \eref{eq:lin_S2_r3} and \eref{eq:lin_R01_r3} for the third region
of \eref{eq:entropy_lin}.
\end{proof}

\section{Summary and Conclusions}

Tight bounds on the entropy of patterns of i.i.d.\ sequences were used to provide
asymptotic and non-asymptotic approximations of the pattern block entropies for
several distributions.  The finite block pattern entropy was approximated for blocks of data generated
by uniform distributions and monotonic distributions.  Monotonic distributions studied
include slowly decaying distributions over the integers, the Zipf distribution, the geometric
distribution, and a linearly increasing distribution.
Specifically, the pattern entropy was bounded for distributions that have infinite i.i.d.\ entropy
rates.  Conditional next index entropy was studied for distributions over small alphabets.

\end{document}